\documentclass{aa}  
\usepackage{graphicx}
\usepackage{txfonts}
\usepackage{tabularx}
\usepackage{amsfonts}
\usepackage{bbold}
\usepackage{color}
\usepackage{transparent}
\usepackage{hyperref}
\usepackage{transparent}
\usepackage{rotating}
\usepackage{caption}
\usepackage{subcaption}

\usepackage{graphicx}	
\usepackage{amsmath}	
\usepackage{amssymb}	
\usepackage{tablefootnote}
\usepackage[flushleft]{threeparttable}
\usepackage{dblfloatfix} 

\usepackage{physics}

\usepackage{natbib}
\bibpunct{(}{)}{;}{a}{}{,} 

\newcommand{\sgx}{SgXB\xspace}

\newcommand*{\hmxb}{HMXB\@\xspace}
\newcommand*{\lmxb}{LMXB\@\xspace}
\newcommand*{\rlof}{RLOF\@\xspace}
\newcommand*{\ns}{NS\@\xspace}
\newcommand*{\eg}{e.g.\@\xspace}
\newcommand*{\ie}{i.e.\@\xspace}

\newcommand{\mystar}{{\fontfamily{lmr}\selectfont$\star$}}
\newcommand*{\msun}{$M_{\odot}$\@\xspace}

\begin{document}

   \title{Formation of wind-captured discs in Supergiant X-ray binaries}

   \subtitle{Consequences for Vela X-1 and Cygnus X-1}

   \author{I. El Mellah
          \inst{1}
          \and
          A. A. C. Sander
          \inst{2,3}
          \and
          J. O. Sundqvist
          \inst{4}
          \and
          R. Keppens
          \inst{1}
          }

   \institute{Centre for mathematical Plasma Astrophysics, 
   			 Department of Mathematics, KU Leuven, 
   			 Celestijnenlaan 200B, B-3001 Leuven, Belgium\\
              \email{ileyk.elmellah@kuleuven.be}
         \and
         	Armagh Observatory and Planetarium, 
         	College Hill, 
         	Armagh, BT61 9DG, Northern Ireland
         \and
             Institut f{\"u}r Physik und Astronomie, 
             Universit{\"a}t Potsdam, 
             Karl-Liebknecht-Str. 24/25, 14476 Potsdam, Germany
         \and
             KU Leuven, Instituut voor Sterrenkunde, 
             Celestijnenlaan 200D, B-3001 Leuven, Belgium
             }

   \date{Received ...; accepted ...}

 
  \abstract
   {In Supergiant X-ray binaries (\sgx), a compact object captures a fraction of the wind of an O/B supergiant on a close orbit. Proxies exist to evaluate the efficiency of mass and angular momentum accretion but they depend so dramatically on the wind speed that given the current uncertainties, they only set loose constrains. Furthermore, they often bypass the impact of orbital and shock effects on the flow structure.
}
   {We study the wind dynamics and the angular momentum gained as the flow is accreted. We identify the conditions for the formation of a disc-like structure around the accretor and the observational consequences for \sgx. 
}
   {We use recent results on the wind launching mechanism to compute 3D streamlines, accounting for the gravitational and X-ray ionizing influence of the compact companion on the wind. Once the flow enters the Roche lobe of the accretor, we solve the hydrodynamics equations with cooling.}
   {A shocked region forms around the accretor as the flow is beamed. For wind speeds of the order of the orbital speed, the shock is highly asymmetric compared to the axisymmetric bow shock obtained for a purely planar homogeneous flow. With net radiative cooling, the flow always circularizes for wind speeds low enough.
}
   {Although the donor star does not fill its Roche lobe, the wind can be significantly beamed and bent by the orbital effects. The net angular momentum of the accreted flow is then sufficient to form a persistent disc-like structure. This mechanism could explain the proposed limited outer extension of the accretion disc in Cygnus X-1 and suggests the presence of a disc at the outer rim of the neutron star magnetosphere in Vela X-1, with dramatic consequences on the spinning up of the accretor.
}


   \maketitle
%

\section{Introduction}

Most stars are found in multiple stellar systems, especially the high mass ones \citep{Duchene2013}. Among them, a significant fraction will undergo a phase of mass transfer which can seriously alter their subsequent evolution \citep{Sana2012}. New observational insights on the long \citep{Abbott2016a} and short term \citep{Grinberg2017} evolution of High Mass X-ray Binaries (\hmxb) have resulted in a compelling need for a more comprehensive description of mass transfer via wind accretion. 


In Supergiant X-ray binaries (\sgx), a supergiant O/B donor star is orbited by a compact object, often a neutron star (\ns), embedded in the stellar wind \citep[for a recent review, see][]{Martinez-Nunez2017}. O/B stars are known to loose mass at a rate up to several 10$^{-6}$M$_{\odot}\cdot$yr$^{-1}$ through a wind whose launching mechanism was first analyzed in detail by \cite{Lucy1970} and \cite{Castor1975} : the resonant line absorption and scattering of UV photons by partly ionized metal ions provides the outer layers of the star with a net outwards momentum. As the flow accelerates, it keeps absorbing previously untouched Doppler-shifted line photons and eventually reaches terminal speeds up to 2,000km$\cdot$s$^{-1}$. It is the gravitational capture of a fraction of this abundant line-driven wind by the compact companion which produces the X-ray luminosity we observe in \sgx, of the order of 10$^{35-37}$erg$\cdot$s$^{-1}$ (\citealt{Walter15}, \citealt{Martinez-Nunez2017} and \citealt{Furst2018} submitted).

Until now, the mass and angular momentum accretion rates pertaining wind accretion have been evaluated based on the Bondi-Hoyle-Lyttleton model \citep[BHL, see][for a review]{Edgar:2004ip} : a planar supersonic flow is gravitationally deflected by the gravitational field of a point-mass and an overdense tail is formed in its wake. The mass accretion rate turns out to be extremely sensitive to the relative speed of the flow with respect to the accretor. In \sgx, the terminal wind speed is generally measured within $\sim$20\% but the accretor lies very close from the stellar surface, in a region where the wind is still accelerating and where orbital effects may significantly alter the picture of a purely radial wind : the theoretical uncertainty on the magnitude and orientation of the wind velocity field within the orbital separation makes the sharp dependency of the BHL mass accretion rate on the wind speed even more crippling. Furthermore, the axisymmetry of the BHL problem has circumvented any discussion on the accretion of angular momentum. This assumption was first relaxed by \cite{Illarionov1975} and \cite{Shapiro1976} to assess the possibility of the formation of a wind-captured disc around compact accretors : they concluded that it was likelier for close binaries, where the star gets close to fill its Roche lobe, but that it was, once again, highly dependent on the relative wind speed. More recently, wind-captured discs have also been identified in simulation of black holes accreting matter from a smooth wind \citep{Walder2014}. Once we complement the models by accounting for the inhomogeneities known to form in this type of winds \citep{Sundqvist2017}, any realistic \sgx X-ray accretion luminosity can be reproduced if we rely only on the BHL formula.

The archetype of a classic \sgx is Vela X-1 where a \ns is on a $\sim$9 days eclipsing orbit around HD 77581, a B0.5 Ib Sg \citep{Hiltner1972,Forman1973}. The \ns is deeply embedded in the intense stellar wind \citep[with a mass loss rate $\dot{M}_{\text{\mystar}}\sim$6.3$\cdot$10$^{-7}$M$_{\odot}$yr$^{-1}$,][]{Gimenez-Garcia2016} with an orbital separation of approximately 1.8 stellar radii \citep{Quaintrell2003a}. The most recent observations revealed a terminal wind speed lower than initially claimed, of the order of 600 to 700km$\cdot$s$^{-1}$, consistent with numerical computation from first principles \citep{Sander2017}. These results suggest that orbital effects might dominate the dynamics between the stellar surface and the \ns, and supply the wind with a significant amount of net angular momentum. It could lead to a complex accretion geometry, with a wind so beamed towards the accretor that it could share some features with the Roche lobe overflow mass transfer (\rlof) at stake in Low Mass X-ray Binaries (\lmxb). On the other hand, we know that in Cygnus X-1, a \sgx hosting a black hole, the O Sg donor star does not fill its Roche lobe \citep{Orosz2011} ; and yet, in spite of the mass being transferred via the stellar wind, the accretor is surrounded by an accretion disc. These two systems suggest that the wind-\rlof regime \citep[first studied in the context of symbiotic binaries by][]{Mohamed2007}, could be the appropriate framework to understand the structure of the accretion flow, rather than the BHL or the \rlof geometries.

A consistent treatment of both the wind acceleration and its accretion by the compact object is thus needed to avoid being left with the wind speed in the vicinity of the accretor as a convenient but unconstraining degree of freedom. \cite{Sander2017} computed the steady state wind stratification for a 1D radial non-local thermal equilibrium atmosphere of a star representative of the donor star in Vela X-1. They accounted for a plethora of chemical elements and ionization levels susceptible to absorb the stellar UV photons, and for the X-ray ionizing feedback from the accretor on the wind ionization state. In this paper, we intend to use this computed 1D line-driven acceleration to see how the 3D structure of the flow departs from a spherical wind once the orbital effects are added. Rather than being based on an empirical fitting formula, the wind velocity and density in the accretion region surrounding the accretor are mere consequences of the stellar and orbital properties. In section\,\ref{sec:orb_dev}, we evaluate the systematic bending of the wind streamlines by the orbital effects, as the wind develops and reaches the Roche lobe of the accretor with a non-zero net angular momentum. Within the Roche lobe, we run 3D hydrodynamical simulations described in section\,\ref{sec:wind-capt_discs} to capture the structure of the flow as it cools down downstream the shock and its capacity to form a disc-like structure. In section\,\ref{sec:obs_cons}, the implications of such a component are discussed in the context of the \sgx Vela X-1 and Cygnus X-1. 

\section{Orbital deviation of the wind}
\label{sec:orb_dev}

\subsection{Model and numerical method}
\label{sec:orb_model}


Sophisticated models and simulations of the launching of line-driven winds show that they become supersonic shortly above the stellar photosphere. This motivates a ballistic treatment of the wind bulk motion at the orbital scale : the trajectory of test-masses is integrated assuming the star and the accretor are on circular orbits and that stellar rotation is synchronized with the orbital period. The 3D equation of motion in the co-rotating frame is :
\begin{equation}
\label{eq:ball}
\boldsymbol{\varv}\dv{\boldsymbol{\varv}}{\boldsymbol{r}} = \boldsymbol{a}_{\text{\mystar}} + \boldsymbol{a}_{\bullet} + \boldsymbol{a}_{\text{ni}}
\end{equation}
where $\boldsymbol{a}_{\bullet}$ stands for the acceleration due to the \ns gravitational field and $\boldsymbol{a}_{\text{ni}}$ for the non-inertial acceleration (centrifugal and Coriolis). The effective acceleration linked to the donor star of mass $M_{\text{\mystar}}$, once projected on the radial unity vector of the spherical frame of the star, is given by :
\begin{equation}
a_{\text{\mystar}}=-GM_{\text{\mystar}}/r_{\text{\mystar}}^2+a_{\text{rad}}\left(r_{\text{\mystar}}\right)+a_{\text{press}}\left(r_{\text{\mystar}}\right)
\end{equation}
where $r_{\text{\mystar}}$ is the distance to the stellar center and $a_{\text{press}}$ is the acceleration due to thermal and turbulent pressure, important near the stellar photosphere. To describe $a_{\text{press}}$ and the total radiative acceleration $a_{\text{rad}}$, with both the line and total continuum contribution, we rely on the computation by \cite{Sander2017} for Vela X-1. Using the stellar atmosphere code PoWR \citep{Hamann1998,Grafener2002}, they calculate an atmosphere model for the donor star assuming a spherical, stationary wind situation. The radiative transfer is performed in the comoving frame, allowing to obtain the radiative acceleration without any further assumption or parametrization :
\begin{equation}
\label{eq:araddef}
a_{\text{rad}}\left(r_{\text{\mystar}}\right)  = \frac{4\pi}{c} \frac{1}{\rho(r_{\text{\mystar}})}  \int\limits_{0}^{\infty} \chi_\nu H_\nu \mathrm{d}\nu 
\end{equation}
with $c$ the speed of light and $\rho$ the mass density, deduced from the stellar mass loss rate and the velocity using the conservation of mass. $\chi_{\nu}$ and $H_{\nu}$ denote the extinction coefficient (in cm$^{-1}$) and the Eddington flux at the frequency $\nu$. Using the technique described in \cite{Sander2017b}, the model provides a hydrodynamically consistent stratification, meaning that the mass loss rate and the velocity field were iteratively updated such that eventually, the outward and inward forces balance each other throughout the stellar atmosphere. The wind velocity profile is expected to follow the $\beta$-law which depends on the terminal speed $\varv_{\infty}$ and on a $\beta$ exponent which quantifies how quickly the terminal speed is reached :
\begin{equation}
\varv_{\beta}(r)=\varv_{\infty}\left(1-R_{\text{\mystar}}/r\right)^{\beta}
\end{equation}
However, the velocity and density stratification computed show notable deviations from this law, especially within a couple of stellar radii where the obtained wind velocity is lower than previously estimated \citep[see the radial velocity profiles in Figure 5 of][]{Sander2017}. In \sgx, it is precisely in this region where the orbiting accretor lies, hence the need to evaluate the impact of the presence of the orbiting accretor on the structure of a slow wind. Notice that in spite of the non-spherical situation due to the presence of the \ns, we adopt $a_\text{rad}$ as $a_\text{press}$ as functions of the distance $r_{\text{\mystar}}$ to the donor star here for the sake of simplicity. It means that we neglect the feedback of the non-radial trajectories on the line acceleration magnitude and orientation.

\begin{figure}
\begin{subfigure}{.5\textwidth}
\centering
\includegraphics[width=0.99\columnwidth]{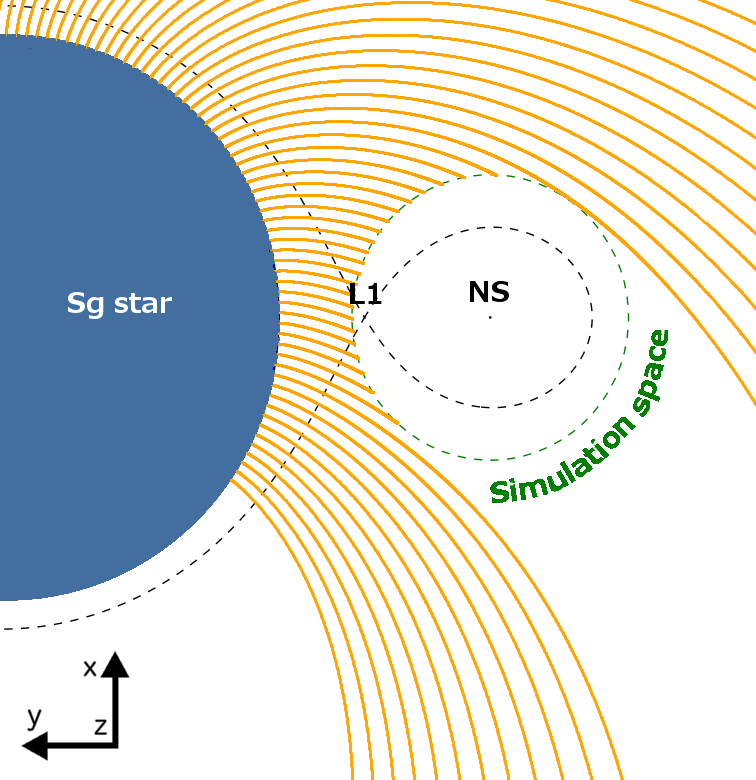}
  \label{fig:sfig1}
\end{subfigure}
\phantom{p}\\
\begin{subfigure}{.5\textwidth}
\centering
\includegraphics[width=0.99\columnwidth]{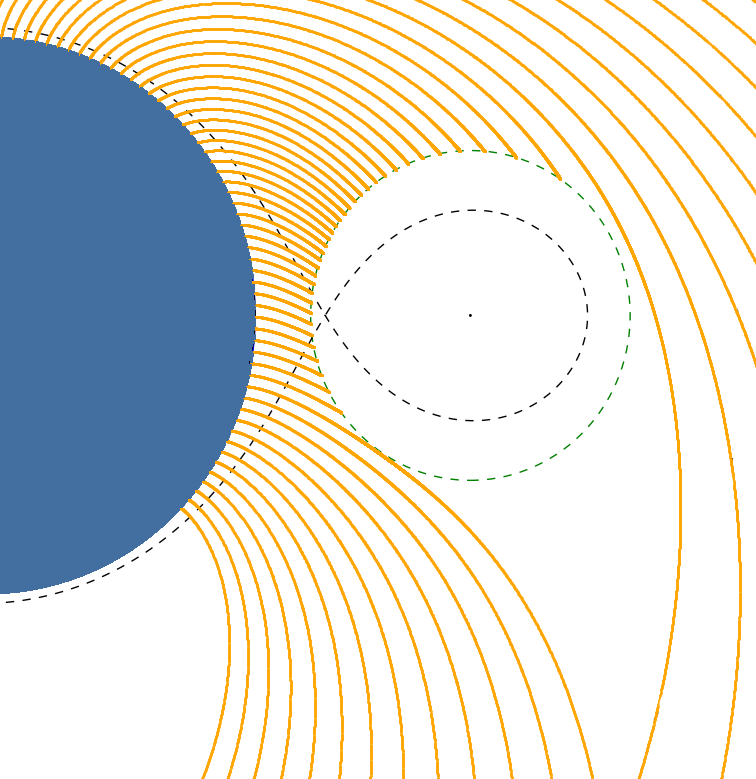}
  \label{fig:sfig2}
\end{subfigure}
\caption{In the orbital plane of the co-rotating frame, a few computed streamlines (orange) from the blue Sg to the HD simulation space (green dashed circle), centered on the accreting NS. The black dashed lines represent the critical Roche surface passing by the first Lagrangian point (L$_1$). Upper panel (resp. lower) is for the light fast (resp. heavy slow) configuration.}
\label{fig:big_picture}
\end{figure}

The streamlines computation is performed using the code developed in \cite{ElMellah2016a}, starting from the stellar surface whose ellipsoidal deformation, even for Roche lobe filling factors close to unity, is not included since it is expected to have a negligible impact on the formation of a wind-captured disc. An illustration of the result is given in Figure\,\ref{fig:big_picture} where the streamlines have been represented in the orbital plane. We stop the integration when the test-masses reach a sphere around the accretor $\sim$30\% larger than its Roche lobe radius. This strategy alleviates the difficulty of an a priori estimate of the accretion radius \citep[the critical impact parameter below which test-masses are captured in the BHL formalism,][]{Edgar:2004ip}. It delimits the space where the ballistic approximation no longer holds. Dissipative effects at shocks will be accounted for within this region in section\,\ref{sec:wind-capt_discs}. With this procedure, we focus on the fraction of the flow susceptible to be eventually accreted rather than on an accurate representation of the accretion tail in the wake of the accretor \citep[for this component, see rather][]{Manousakis2014}.

\begin{figure*}[!b]
\centering
\includegraphics[width=2\columnwidth]{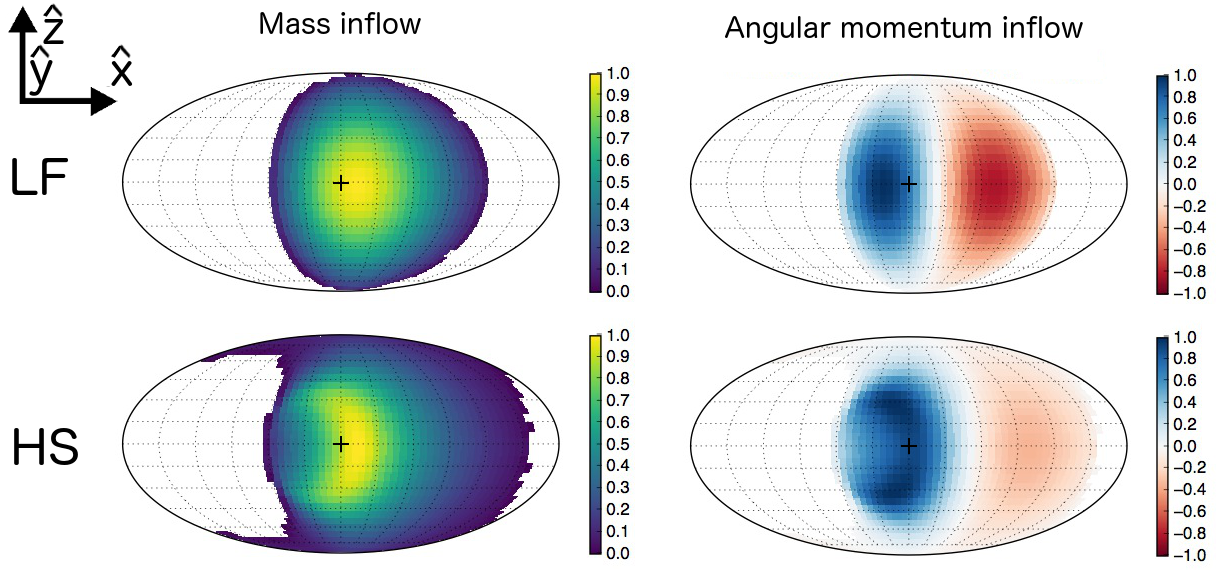}
\caption{Mollweide projections of local mass and angular momentum inflows within the simulation space centered on the accretor (dashed green sphere on Figure\ref{fig:big_picture}). The upper row corresponds to the light fast (LF) case while the bottom row is for the heavy slow (HS) case. Each map is scaled to its maximum (absolute) value and centered on the axis from the accretor to the donor star (central dark cross). No mass inflow is measured in the white angular regions of the left maps (\ie they are not reached by any of the ballistic streamlines). Positive (resp. negative) values of angular momentum stands for locally prograde (resp. retrograde) flow with respect to the orbital motion.}
\label{fig:inflow_maps}
\end{figure*} 

The wind terminal speed is expected to scale approximately as the effective escape velocity (\ie once surface gravity has been corrected for radiative continuum pressure on free electrons via the Eddington parameter). \cite{Vink2001} showed that for stellar effective temperatures above $\sim$25kK, the proportionality constant is twice higher than below, leading to a sharp drop of the terminal wind speed as we go towards lower temperatures. The donor star in Vela X-1, HD 77581, is a B0.5 Ib Sg star \citep{Hiltner1972,Forman1973} whose effective temperature is $\sim$25kK. \cite{Gimenez-Garcia2016} suggested that it could explain the low terminal speed of 700km$\cdot$s$^{-1}\pm$100km$\cdot$s$^{-1}$ they measured for the wind of HD 77581. The computation carried out by \cite{Sander2017} for HD 77581 also leads to terminal speeds ranging from 400 to 600km$\cdot$s$^{-1}$ depending on the inclusion of X-ray illumination from the accretor. A decisive result of their analysis is that the latter modifies the ionization state of the metal ions in the wind but does not necessarily inhibit the acceleration process. On the contrary, far enough upstream the \ns, the effective absorption of UV photons might be locally enhanced once the metal ions are in a higher ionization level. It is only close from the accretor, once all the elements have been deprived of their electrons, that the line-driven acceleration is halted, as previously emphasized in the literature \citep[see \eg][]{Hatchett1977,Ho1987,Blondin1990a,Karino2014}. In an attempt to illustrate the dramatic impact of the efficiency of the line-driven acceleration on the subsequent properties of the accretion flow, and to encompass potential inaccuracies in the calculation of this acceleration in the context of Vela X-1, we consider the case of an artificially enhanced wind acceleration (by 50\%) which leads to larger flow velocities by approximately 20\%. It could be caused, for instance, by a different chemical composition of the stellar atmosphere than the one assumed in \cite{Sander2017}. In section\,\ref{sec:wind-capt_discs}, we will see that the orbital speed is a threshold which separates two types of accretion flows and given the value of the orbital speed in Vela X-1 ($\sim$284km$\cdot$s$^{-1}$), this enhancement of the wind acceleration will induce major changes. From now on, we consider the two cases in Table\,\ref{tab:params} :
\begin{itemize}
\item \underline{the heavy slow (HS) :} the accretor is heavy, with a mass of M$_{\bullet}$=2.5\msun, lying on the upper edge of the expected maximum mass for a \ns, and the radiative acceleration efficiency is not enhanced, leading to a relatively slow wind.
\item  \underline{the light fast (LF) :} the accretor has a mass of M$_{\bullet}$=1.5\msun and the radiative acceleration efficiency is enhanced by 50\%.
\end{itemize} 
Since the \ns mass estimates in Vela X-1 range from 1.7\msun \citep{Rawls2011} up to 2.3\msun \citep{Quaintrell2003a}, partly due to the uncertainty on the inclination of the system, we expect the real configuration to lie in-between the two cases we consider.

\subsection{Inhomogeneity and asymmetry of the wind}
\label{sec:orb_inhomo}

We now monitor the asymmetry and inhomogeneity of the flow when it reaches the spherical HD simulation space centered on the compact object and corresponding approximately to its Roche lobe. The aforementioned ballistic integration supplied information on the velocity vector at the surface of this sphere while the density relative to the one at the stellar photosphere is deduced from the divergence of each streamline. This information is then binned on angular tiles, with the polar axis of the spherical frame aligned with the orbital angular momentum axis ($\hat{z}$ in Figure\,\ref{fig:big_picture} and \ref{fig:inflow_maps}). In Figure\,\ref{fig:inflow_maps}, we represented Mollweide projections of the local mass and angular momentum inflow at the surface of this space for the HS and LF cases : it offers an overview of the properties of the flow entering the accretor Roche lobe, seen from the accretor. 

Concerning the integrated values at the inflowing edge of the simulation space, we focus on the mass inflow rate $\dot{M}_{\text{out}}$, the net specific (\ie per unit mass) angular momentum of the flow $l_{\text{out}}$ and its corresponding circularization radius $R_{\text{circ}}$. The circularization radius is the radius at which a circular Keplerian orbit around the accretor would have the same specific angular momentum \ie $R_{\text{circ}}=l_{\text{out}}^2/GM_{\bullet}$. The values are given in Table\,\ref{tab:params} and compared respectively to the stellar mass loss rate $\dot{\text{M}}_{\text{\mystar}}$, to the orbital specific angular momentum $a^2\Omega$ and to the \ns magnetosphere radius R$_{\text{mag}}$ \citep[see \eg][]{Frank2002} :
\begin{align}
\begin{split}
\label{eq:Rmag}
R_{\text{mag}}\sim & 1.4\cdot 10^9\text{cm}\left(\frac{\rho}{10^{-12}\text{g}\cdot\text{cm}^{-3}}\right)^{-1/6}\left(\frac{\varv}{2,000\text{km}\cdot\text{s}^{-1}}\right)^{-1/3} \text{...}\\
& \text{...} \quad \left(\frac{B_{\bullet}}{2.6\cdot 10^{12}\text{G}}\right)^{1/3}\left(\frac{R_{\bullet}}{10\text{km}}\right)
\end{split}
\end{align}
where the values used for the mass density $\rho$ and the flow speed $\varv$ are orders-of-magnitude at the outer edge of the magnetosphere. The low dependence of the magnetosphere radius on these parameters guarantees that their exact value will not significantly alter this estimate. A typical \ns radius has been used and the \ns magnetic field is the one deduced by \cite{Furst2014} in Vela X-1. In Table\,\ref{tab:params}, we used R$_{\text{mag}}=$1.4$\cdot$10$^9$cm. We expect any disc-like structure to be truncated approximately at the inner radius \citep{Ghosh1978} while quasi-spherical accretion onto the magnetosphere would proceed as described by \cite{Shakura2013b}. Notice that the mass inflow rates displayed in Table\,\ref{tab:params} set only upper limits on the final rate at which matter will be accreted since only a subset of the streamlines entering the Roche lobe of the accretor will eventually be accreted. However, it is already striking to notice how much more important the fraction of the stellar wind entering the simulation space in the HS configuration is compared to the LF one. The difference is essentially due to a significantly more important contribution of the high latitude region in the former case. Concerning the angular momentum, it might still vary within the simulation space since the forces are not isotropic around the accretor. The results displayed in Table\,\ref{tab:params} show that. Within the current uncertainties on the mass of the accretor and on the efficiency of the wind launching process in Vela X-1, the 2 cases lead to dramatically different accretion flow configurations, in spite of their apparently similar parameters. In the HS case, where the flow is slightly slower than the orbital speed, the mass inflow rate within the Roche lobe of the accretor is 4 times larger, while the circularization radius is almost an order of magnitude larger than in the LF case, where the flow is slightly faster than the orbital speed. 

In the left panels in Figure\,\ref{fig:inflow_maps}, we see that the bulk of the mass inflow is approximately distributed in the same way in both cases, with a larger off-plane contribution when the wind is slower : it is a first hint that the inertia of the wind is no longer large enough to overcome the orbital flattening induced by rotation, a feature which will have major consequences within the shocked region. In both cases, the incoming flow is centered around a mean radial direction which departs from the axis joining the compact object to the star (central cross in Figure\,\ref{fig:inflow_maps}). The main difference though lies in the distribution of angular momentum inflow (right panels) : the LF case leads to an equivalent amount of positive and negative angular momentum, which shows the essentially planar (albeit deviated) structure of the flow, whereas the HS case displays a large unbalance. The evaluation of the net angular momentum inflowing was in no case obvious a priori : the flow arriving from the first Lagrangian point L$_1$ (with positive angular momentum) is denser than the flow arriving from the right of L$_1$ as seen from the accretor (with negative angular momentum), but it is also slower. The present analysis shows that the former effect eventually dominates. The non-zero net angular momentum in the HS case can not be attributed to an asymmetry of the mass inflow. Rather, it is due to the shift between the mean direction of arrival of matter (yellow spot in mass inflow maps) and the direction of radial inflow (white stripe in-between blue and red in angular momentum inflow maps). It is much more significant for HS than for LF. Consequently, the net amount of specific angular momentum is larger for HS, which also leads to larger circularization radii and to the likelier presence of a wind-captured disc, a prediction we now put to the test.
\begin{center}
\begin{table}[!t]
\caption{Parameters representative of Vela X-1 (with the star indexed with \mystar\xspace and the \ns with $\bullet$), described in the text. The bottom part of the table displays integrated quantities at the outer edge of the simulation space for the 2 models considered (LF and HS).}
\label{tab:params}
\centering
\begin{tabularx}{0.53\columnwidth}{c|c|c}
   & LF & HS \\
  \hline
  M$_{\text{\mystar}}$ & \multicolumn{2}{c}{20.2M$_{\odot}$} \\
  R$_{\text{\mystar}}$ & \multicolumn{2}{c}{28.4R$_{\odot}$} \\
  P$=2\pi/\Omega$ & \multicolumn{2}{c}{8.964357 days} \\  
  a$/$R$_{\text{\mystar}}$ & \multicolumn{2}{c}{$\sim$1.8}\\
  $\dot{\text{M}}_{\text{\mystar}}$ & \multicolumn{2}{c}{6.3$\cdot$10$^{-7}$M$_{\odot}\cdot$yr$^{-1}$} \\
  \hline
  M$_{\bullet}$ & 1.5M$_{\odot}$  & 2.5M$_{\odot}$  \\
  Enhanced & Yes & No  \\
  \hline
  $\dot{\text{M}}_{\text{out}}/\dot{\text{M}}_{\text{\mystar}}$ & 4\% & 17\% \\
  $l_{\text{out}}/a^2\Omega$ & -1\% & 3\% \\
  R$_{\text{circ}}$ / R$_{\text{mag}}$ & 4 & 30 \\
\end{tabularx}
\end{table}
\end{center}

\section{Wind-captured discs}
\label{sec:wind-capt_discs}

\subsection{Physics and numerical setup}
\label{sec:HD}

\subsubsection{Equations}
\label{sec:HD_eq}

Within the Roche lobe of the accretor, we solve the non-viscous HD equations in their conservative form, converting accordingly the gravitational, radiative and non-inertial accelerations in the ballistic equation of motion\,\eqref{eq:ball} into forces per unit volume. In a first model, we do solve the adiabatic energy equation everywhere, assuming that the heating from the donor star and from the X-rays produced in the vicinity of the accretor balances the cooling of this optically thin supersonic wind. In section\,\ref{sec:cool}, we will discuss the validity of this adiabatic approximation downstream the shock which will form, and a way to relax it. As is custom, we close the system of equations by assuming an ideal gas with an adiabatic index $\gamma=5/3$

The computation is performed with the finite volume code \texttt{MPI-AMRVAC} \citep{Xia2017}, using a 3$^{\text{rd}}$ order HLL solver \citep{Toro1994} with a Koren slope limiter \citep{Vreugdenhil1993}. The spherical mesh we set up is an extension of what has been developed for an axisymmetric 2D flow in \cite{ElMellah2015} : it is centered on the accretor and radially stretched to guarantee a constant relative resolution from the outer to the inner edge of the simulation space, spanning several orders of magnitude at an affordable computational cost and with a uniform cell aspect ratio. The outer radius of the simulation space is approximately half of the orbital separation, which is $\sim$0.2AU in Vela X-1, while the inner edge has a radius of 5 times the \ns magnetosphere radius given by equation\,\eqref{eq:Rmag}, hence a factor of approximately 200 between the inner and outer edge. Due to the symmetry of the problem above and below the orbital plane, we consider only the upper hemisphere and work with a resolution of 128$\times$32$\times$128 corresponding to cells of aspect ratio close to unity near the equatorial plane of the mesh. Since our aim is to identify the conditions suitable for the formation of a wind-captured disc, the conservation of angular momentum is of uttermost importance. We implemented an angular momentum preserving scheme which guarantees the conservation of the component of the angular momentum projected onto the polar axis, in particular in the innermost regions of the flow, to machine precision. Instead of solving for the azimuthal component of the linear momentum, we consider the projection of the equation of conservation of angular momentum along the polar axis of the spherical frame \citep[similarly to what was done in polar coordinates by][]{Molteni1999} :
\begin{equation}
\begin{split}
\partial_t L_z + & \frac{1}{r^2}\partial_r\left(L_z v_{r} r^2 \right) + \frac{1}{r\sin^2\theta}\partial_{\theta}\left(L_z v_{\theta} \sin^2\theta \right) \\
& + \frac{1}{r\sin\theta} \partial_{\phi}\left(L_z v_{\phi}\right) + \partial_{\phi}P=\Sigma
\end{split}
\end{equation}
where $t$ stands for the time coordinate, $r$, $\theta$ and $\phi$ are the classic spherical coordinates in the frame centered on the accretor, $\rho$ is the mass density, $P$ the thermal pressure and $v_{r}$, $v_{\theta}$ and $v_{\phi}$ are respectively the radial, polar and azimuthal components of the velocity. $\Sigma$ is the source term due to the non-isotropy of the potential around the \ns that we do not explicit here. Finally, $L_z=\rho r v_{\phi} \sin\theta$ is the angular momentum per unit volume projected on the polar axis. Since no geometrical source term is left after spatial (and time) discretization, numerical conservation of the polar component $L_z$ of the angular momentum is guaranteed.

\subsubsection{Radiative cooling}
\label{sec:cool}

Let us estimate the importance of cooling in this physical environment. Upstream the shock, we rely on the temperature stratification derived from the solution of the statistical equilibrium equations and the radiative transfer. Based on the assumption of radiative equilibrium, the (electron) temperature structure in the expanding atmosphere is obtained by applying a generalized Uns{\"o}ld-Lucy method described in \citet{Hamann1998}. Including the X-ray irradiation on the donor star leads only to a moderate increase of the temperature profile upstream the accretor compared to what was obtained in \cite{Sander2017}. Downstream the shock, we assume that the gas is optically thin and write the time scale $\tau_{\text{c}}$ to radiate away the internal energy : 
\begin{equation}
\tau_{\text{c}}=\frac{nk_{\text{B}}T}{n^2\Lambda\left(T\right)}
\end{equation}
where $k_{\text{B}}$ is the Boltzmann constant, $T$ is the temperature, $n$ the Hydrogen number density and $\Lambda\left(T\right)$ the cooling rate computed by \cite{Schure2009}, which includes the proportion of electrons relative to protons. If $\tau_{\text{d}}$ is the dynamical time scale for free fall at a fiducial accretion radius R$_{\text{acc}}$ of 1$/$30$^{\text{th}}$ of the orbital separation in Vela X-1, we obtain :
\begin{align}
\begin{split}
\frac{\tau_{\text{c}}}{\tau_{\text{d}}}\sim & 0.01\left(\frac{T}{10^6\text{K}}\right)\left(\frac{\Lambda}{10^{-22}\text{erg}\cdot\text{s}^{-1}\cdot\text{cm}^{3}}\right)^{-1} \text{...}\\
&  \text{...} \quad \left(\frac{\rho}{10^{-13}\text{g}\cdot\text{cm}^{-3}}\right)^{-1}\left(\frac{M_{\bullet}}{2\text{M}_{\odot}}\right)^{1/2}\left(\frac{\text{R}_{\text{acc}}}{0.2\text{AU}/30}\right)^{-3/2}
\end{split}
\end{align}
where we use the values of temperature and density measured downstream the shock in the adiabatic simulations presented in section\,\ref{sec:cool_F}. Except if the wind is a few times faster than expected in Vela X-1 and/or the star displays a mass loss rate an order of magnitude lower than what models and observations indicate, the flow will be dense enough to significantly cool in the shocked region.

In an optically thin environment, we could in principle include radiative cooling using the module developed for \texttt{MPI-AMRVAC} by \cite{VanMarle2011}. Given the aforementioned ratio, it would lead to a cooling of the flow down to a floor temperature set by the X-ray and stellar radiation heating. Also, the optically thin approximation might not hold within the shocked region, especially when runaway cooling occurs and a high density disc-like region forms. As such, we chose to represent the cooling in a simpler way, using a polytropic model. It is equivalent to assume that the ratio of energy radiated away by the work done by the pressure force is constant : a certain compression leads to a certain energy loss, ranging from 0 (in the adiabatic limit) to 100\% (in the isothermal limit) of the work done by the pressure force \citep{Christians2012}. Above a certain threshold temperature T$_{0}$, reached only within the shocked region, we overwrite the solution for the internal energy computed by the energy equation with the corresponding value of pressure deduced from the polytropic relation :
\begin{equation}
P=C\rho^{\alpha}
\end{equation}
Provided there is no creation of entropy (in particular no shock), $C$ is constant and uniform. In this framework, the polytropic index $\alpha$ ranges from 1 in the isothermal limit to $\gamma$ in the adiabatic limit \citep{Horedt2000}. After exploring a range of possible values for $C$ and $\alpha$, we retained 3 different cooling models :
\begin{itemize}
\item \underline{Isentropic (or "isoS") :} cooling occurs only in a thin unresolved radiative layer immediately downstream the shock and is then negligible (for instance, because of intense X-ray heating) which means $\alpha=\gamma$ and a constant $C$ set to a few percent of the entropy the flow would acquire downstream the shock in the fully adiabatic case, $S_0$. T$_{0}$ is set to 10$^6$K.
\item \underline{Isothermal hot (or "hot") :} above $T_0=C=10^6$K, the net cooling is efficient enough to compensate any adiabatic compression as the flow accretes, which leads to an isothermal flow ($\alpha=1$).
\item \underline{Isothermal cool (or "cool") :} same as previous but with a temperature $T_0=10^5$K.
\end{itemize}
In the two isothermal cases, the cooling prescription means that the flow evolution is fully adiabatic until it reaches the temperature $T_0=C$ when it becomes isothermal. We believe that including optically thin cooling without heating would lead to results qualitatively similar to the isothermal prescription we introduce here \citep[as noticed by][]{Saladino2018} . The four models (fully adiabatic, isentropic, isothermal hot and cool) are summarized in Table\,\ref{tab:cool}.

\begin{table}
\centering
\caption{Parameters of cooling prescriptions in the four models.}
\label{tab:cool}
\begin{tabularx}{0.78\columnwidth}{c|c|c|c|c}
   & adiabatic & isoS & Hot & Cool\\
  \hline
  Cooling & no & yes & yes & yes \\  
  $T_0$ & -- & 10$^6$K & 10$^6$K & 10$^5$K \\
  $C$ & -- & $S_0$ & $T_0$ & $T_0$ \\
  $\alpha$ & -- & $\gamma$ & 1 & 1 \\
\end{tabularx}
\end{table}

\subsection{Flow morphology}
\label{sec:morph}

\subsubsection{Adiabatic evolution}
\label{sec:cool_F}

\begin{figure}
\begin{subfigure}{.5\textwidth}
\centering
\includegraphics[width=0.99\columnwidth]{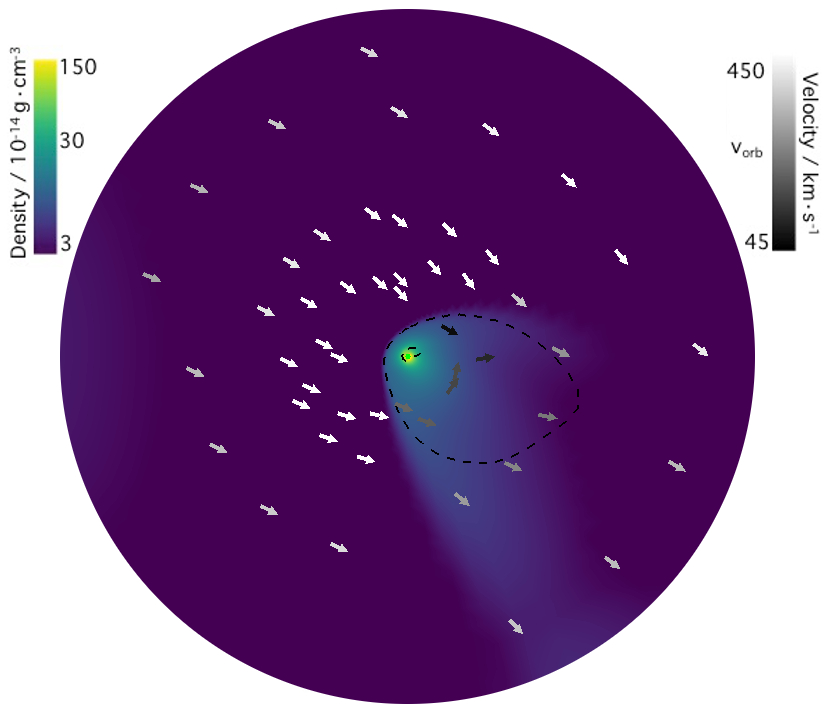}
  \label{fig:sfig1}
\end{subfigure}
\begin{subfigure}{.5\textwidth}
\centering
\includegraphics[width=0.99\columnwidth]{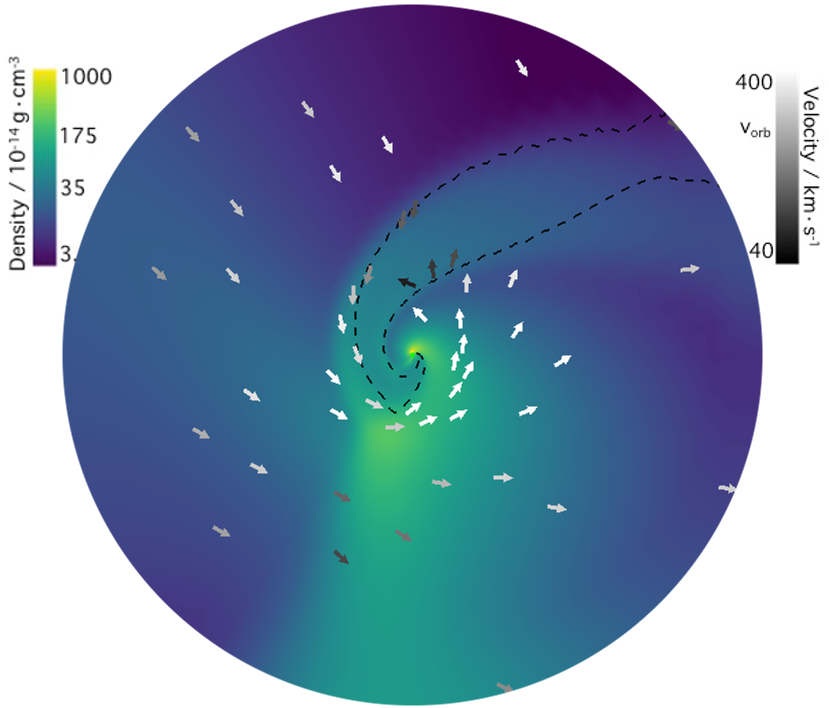}
  \label{fig:sfig2}
\end{subfigure}
\caption{Logarithmic colormaps of the density field in the orbital plane. The arrows stand for the velocity field, with a black to white colormap for their increasing magnitude. The orbital speed of Vela X-1, v$_{\text{orb}}\sim284$km$\cdot$s$^{-1}$, has been represented (linear scale). The black dashed line is the Mach-1 contour. The radial extension of the simulation domain relative to the orbital separation corresponds to the green dashed delimited regions in Figure\,\ref{fig:big_picture}, which is approximately the Roche lobe of the accretor. (bottom) LF configuration. (top) HS configuration.}
\label{fig:adiab}
\end{figure} 

\begin{figure}
\centering
\includegraphics[width=0.99\columnwidth]{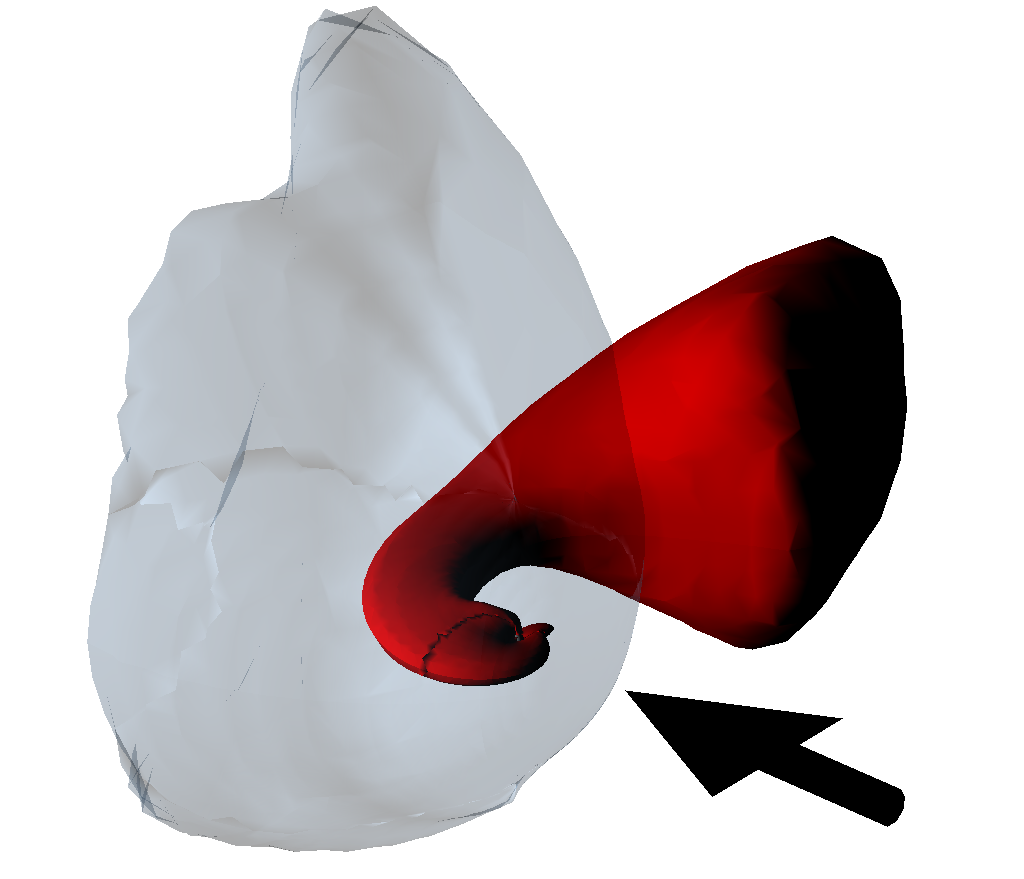}
\caption{3D contours of the mass density for the LF (semi-transparent blue) and HS (red) configurations. The black arrow indicates the approximate direction of the arriving wind, while the vertical direction is aligned with the orbital angular momentum. Notice the axisymmetry of the LF flow structure around the mean direction of wind arrival, whereas the HS flow is compressed in the orbital plane and forms a characteristic channel reminiscent of the stream of matter in \rlof systems. Same scale as Figure\,\ref{fig:adiab}.}
\label{fig:3D_adiab}
\end{figure}

In Figure\,\ref{fig:adiab}, we represent slices in the orbital plane of the numerically relaxed state reached by the simulations. A 3D representation is displayed in Figure\,\ref{fig:3D_adiab} to appreciate the level of beaming of the flow in the orbital plane.

In the case of a light accretor capturing material from a fast wind (LF configuration, upper panel in Figure\,\ref{fig:adiab}), the main features depart little from what has been observed in simulations of axisymmetric uniform flows. In agreement with \cite{Blondin:2012vf}, we do not observe any transverse oscillation of the tail \citep[the so-called "flip-flop instability" which arises mostly in 2D polar numerical setups,][]{Foglizzo2005}. The orbital effects deflect the wind whose mean direction of arrival is $\sim$20 degrees misaligned with respect to the axis joining the star to the compact object. However, as discussed in section\,\ref{sec:orb_inhomo}, the flow remains essentially planar around this direction. When the flow is sufficiently beamed towards the accretor, it forms a bow shock (semi-transparent blue surface in Figure\,\ref{fig:3D_adiab}) at a distance ahead the accretor compatible with a fraction of the accretion radius \citep{Edgar:2004ip}. The Mach-1 surface and the cone of density jump are slightly misaligned with each other, with the side facing the star denser. The Mach number immediately upstream the shock reaches 30 and we retrieve the classic jump conditions for an adiabatic shock. Between the outer boundary upstream and the inner boundary, the density (resp. the temperature) increases by a factor of $\sim$100 (resp. 5,000). In the innermost regions of the flow, we retrieve the sonic surface though no longer anchored into the inner boundary, contrary to what was predicted for a planar uniform flow with $\gamma=5/3$ by \cite{Foglizzo1996}.

In the case of a heavy accretor capturing material from a slow wind (HS configuration, lower panel in Figure\,\ref{fig:adiab}), the morphology of the flow is dramatically different. Not only is the mean direction of arrival of the flow more misaligned with the line joining the star to the compact object ($\sim$45 degrees) but also the shearing is much more important, leading to a significant amount of net angular momentum. A bow shock also forms but while it extends over several accretion radii on the side where the flow is less dense and faster, the beamed wind arriving directly from L$_1$ remains mildly supersonic as it passes the accretor. It is strongly deflected and accelerated by the gravitational slingshot but only to finally impact the shocked region from the back. The adiabatic compression it first experiences leads to a dense and fairly cool region compared to the innermost parts of the flow. We also notice that when the wind is slower, material from higher latitudes on the star contributes to the accretion process, as shown by the vertical extent of the mass inflow map in Figure\,\ref{fig:inflow_maps} : the beaming of the wind in the orbital plane builds up the red dark bulge observed on the right in Figure\,\ref{fig:3D_adiab}. It is a specific feature of wind \rlof configurations since in pure \rlof, only matter from the vicinity of L$_1$ flows in the Roche lobe of the accretor while in pure wind configurations, the centrifugal force is too weak to focus the fast wind in the orbital plane and the wind at high stellar latitudes does not participate in the accretion process. Downstream this bulge, as seen in Figure\,\ref{fig:3D_adiab}, the shocked region of the HS setup presents a characteristic spiral shape which delimits a narrow accretion channel along which matter flows in (or out beyond the stagnation point). The orientation of this stream differs in its orientation with the one observed in \rlof systems due to the much lower effective gravity of the donor star (after including the line radiation and pressure accelerations), which alters the classic Roche potential we rely on in \lmxb.

Although the Mach number of the flow entering the simulation space remains below 10 due to the limited efficiency of the wind acceleration, it reaches Mach numbers of 20 just upstream the shock, leading to a temperature jump of approximately 400. As the flow is accreted, the corresponding temperatures of the order of 10MK keeps increasing up to 100MK at the inner boundary. This temperature, close from the relativistic regime, is unphysical in this context and a mere consequence of the adopted adiabatic treatment ; however, as discussed already in section\,\ref{sec:cool}, a more realistic one would also include effects of radiative cooling in the simulations.

\subsubsection{Polytropic cooling}
\label{sec:cool_T}

\begin{figure*}
\begin{subfigure}{0.5\textwidth}
\begin{center}
\includegraphics[width=8.1cm]{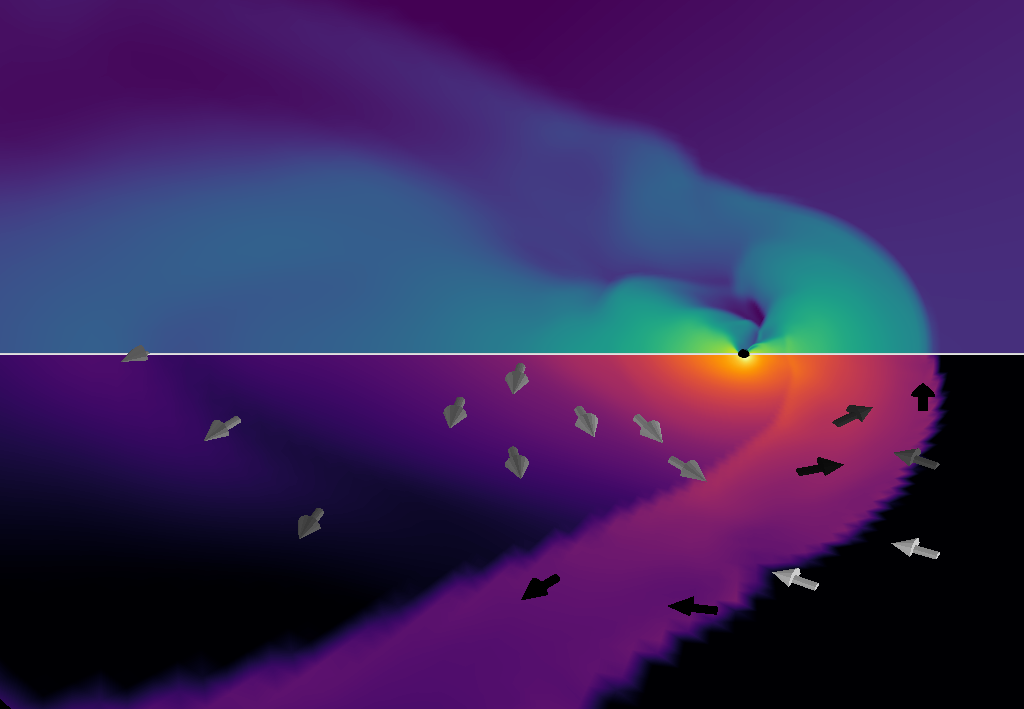} 
\label{fig:subim1}
\end{center}
\end{subfigure}
\begin{subfigure}{0.5\textwidth}
\begin{center}
\includegraphics[width=8.1cm]{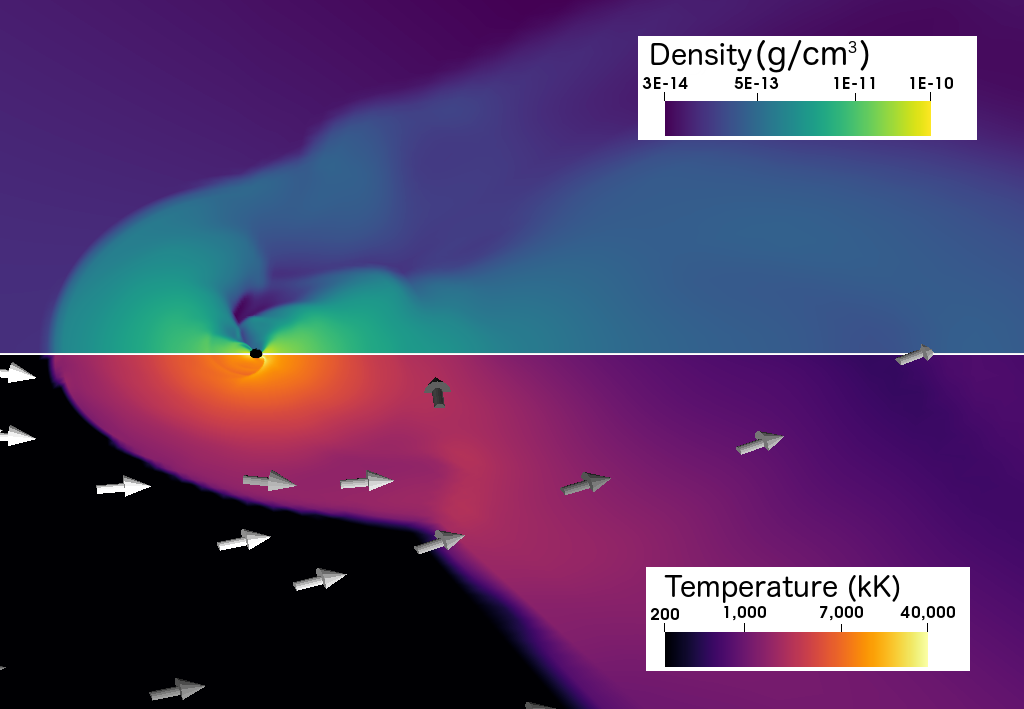}
\label{fig:subim2}
\end{center}
\end{subfigure}
\vspace*{0.8cm}\\
\hspace*{-0.2cm}
\begin{subfigure}{0.5\textwidth}
\begin{center}
\includegraphics[width=8.1cm]{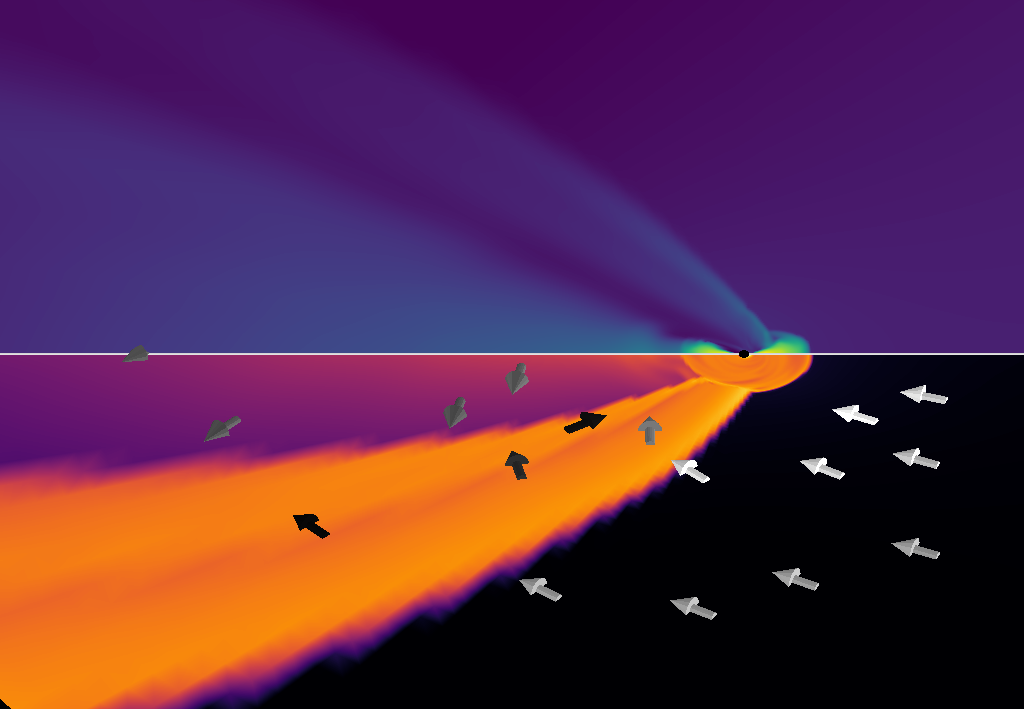}
\label{fig:subim2}
\end{center}
\end{subfigure}
\begin{subfigure}{0.5\textwidth}
\begin{center}
\includegraphics[width=8.1cm]{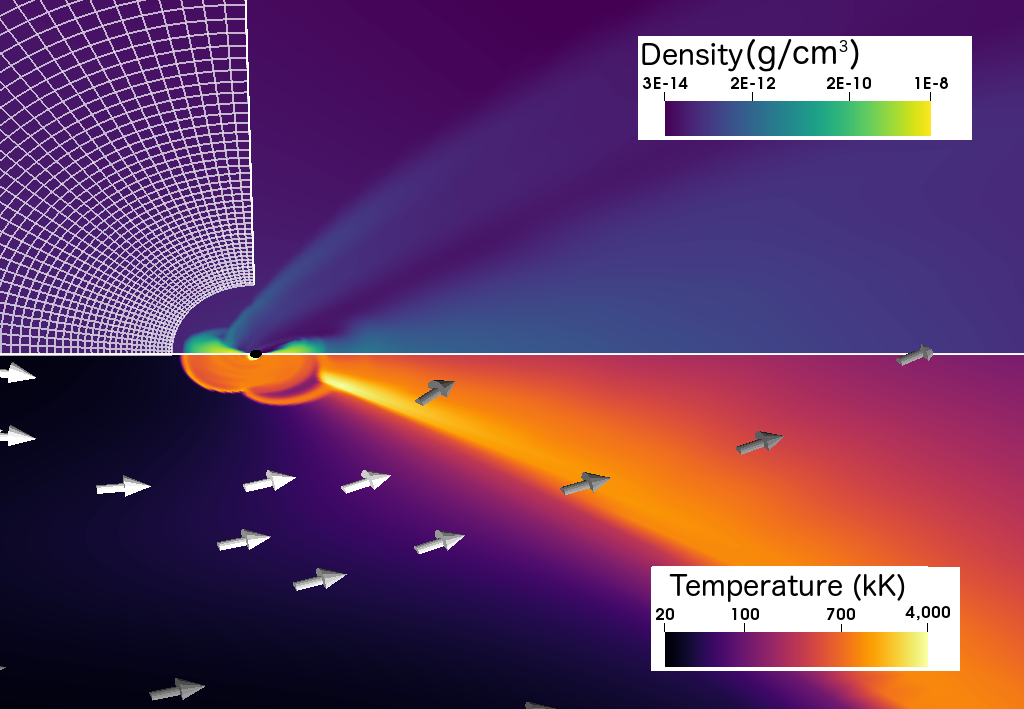} 
\label{fig:subim1}
\end{center}
\end{subfigure}
\vspace*{0.8cm}\\
\hspace*{-0.2cm}
\begin{subfigure}{0.5\textwidth}
\begin{center}
\includegraphics[width=8.1cm]{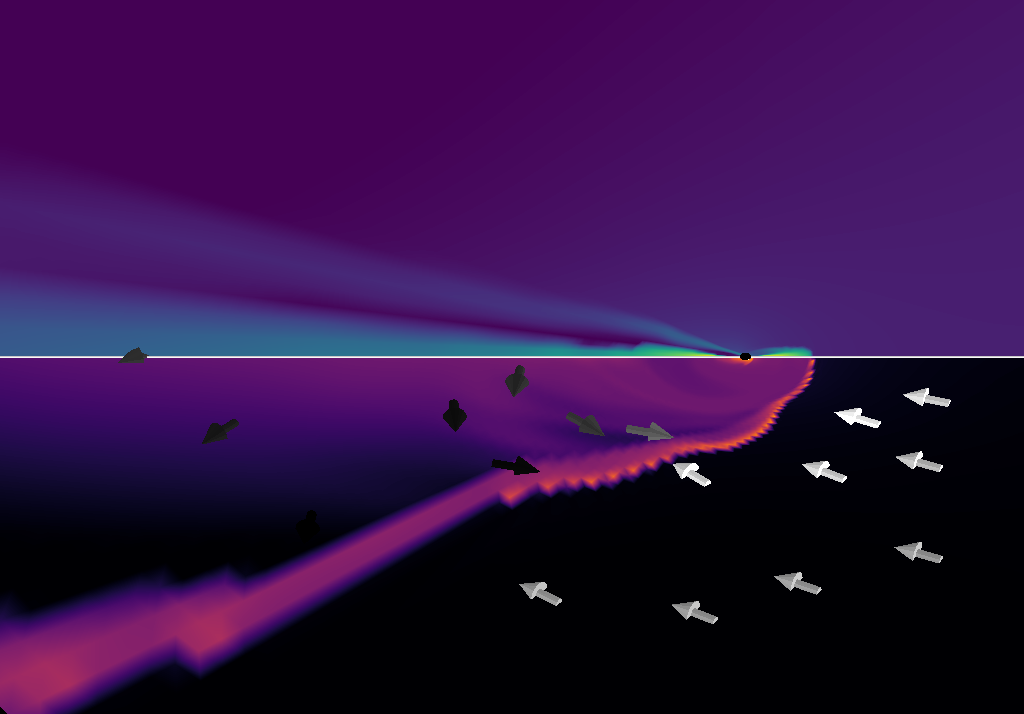}
\label{fig:subim2}
\end{center}
\end{subfigure}
\begin{subfigure}{0.5\textwidth}
\begin{center}
\includegraphics[width=8.1cm]{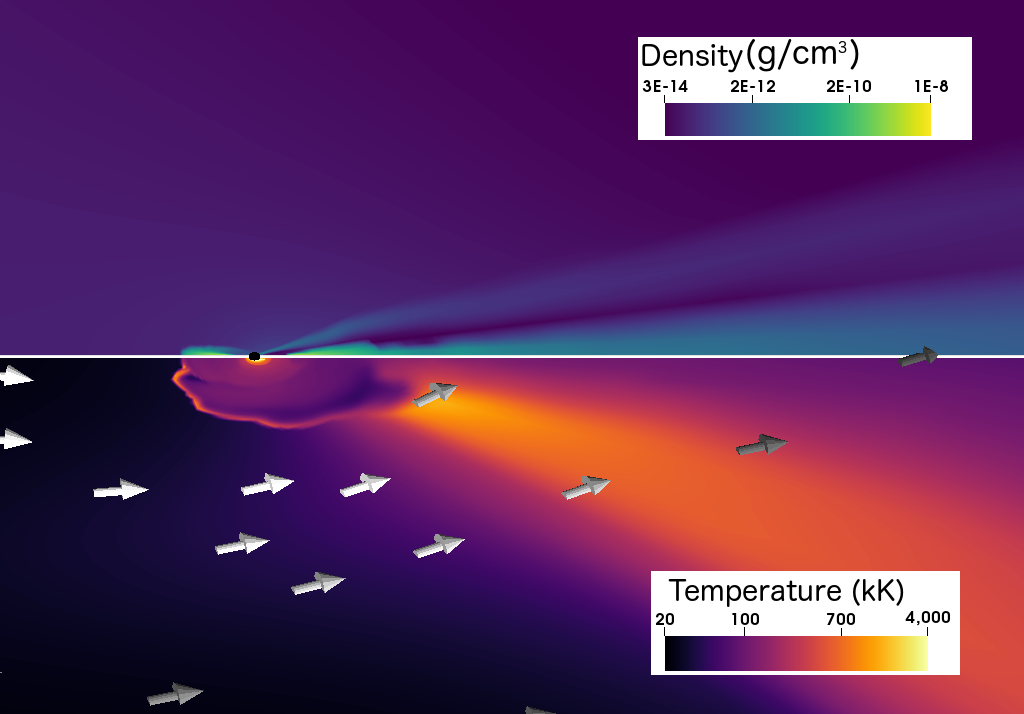}
\label{fig:subim2}
\end{center}
\end{subfigure}
\caption{Side-views of the flow structure when cooling is triggered using an isentropic (upper panels) or an isothermal prescription, with a high temperature (middle) or low temperature (lower panels). In the left (resp. right) column, the wind comes from the right (resp. left). The lower half of each panel displays a logarithmic thermal colormap in the orbital plane while the upper half represents the transverse (or "vertical") logarithmic density distribution. We also plotted the velocity field in the orbital plane, with white to black color scale to indicate a slowing down by a factor of at least 4. The radially stretched mesh has been represented to indicate the resolution.}
\label{fig:cooled_ones}
\end{figure*}

\begin{figure}
\centering
\includegraphics[width=0.95\columnwidth]{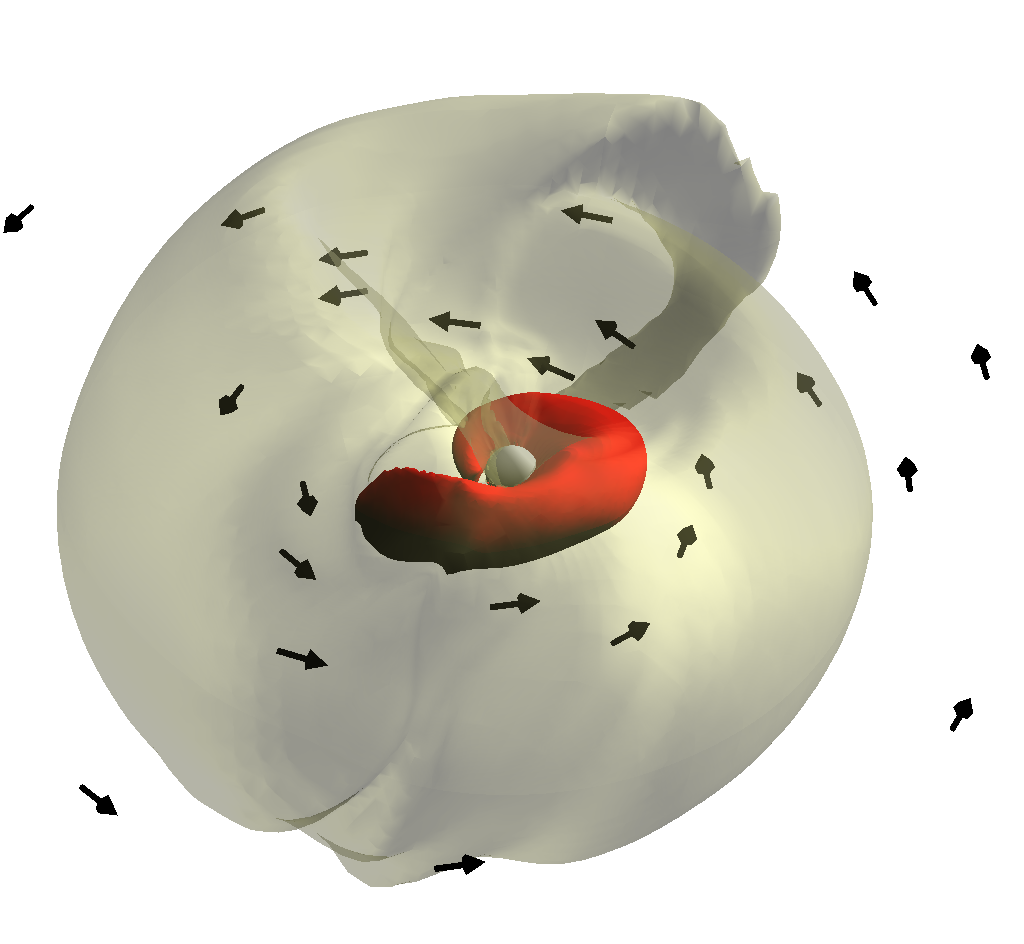}
\caption{3D contours of the mass density for the isentropic HS configuration (upper panels in Figure\,\ref{fig:cooled_ones}), with the yellow semi-transparent surface 5 times less dense than the inner red surface. The arrows stand for the velocity field in the orbital plane. The flow comes from the upper left. Spiral arms are visible for each surface. The central white sphere stands for the inner boundary of the simulation space, $\sim$ 200 times smaller than the outer boundary displayed in Figure\,\ref{fig:big_picture}.}
\label{fig:disc}
\end{figure} 

\begin{figure}
\centering
\includegraphics[width=0.95\columnwidth]{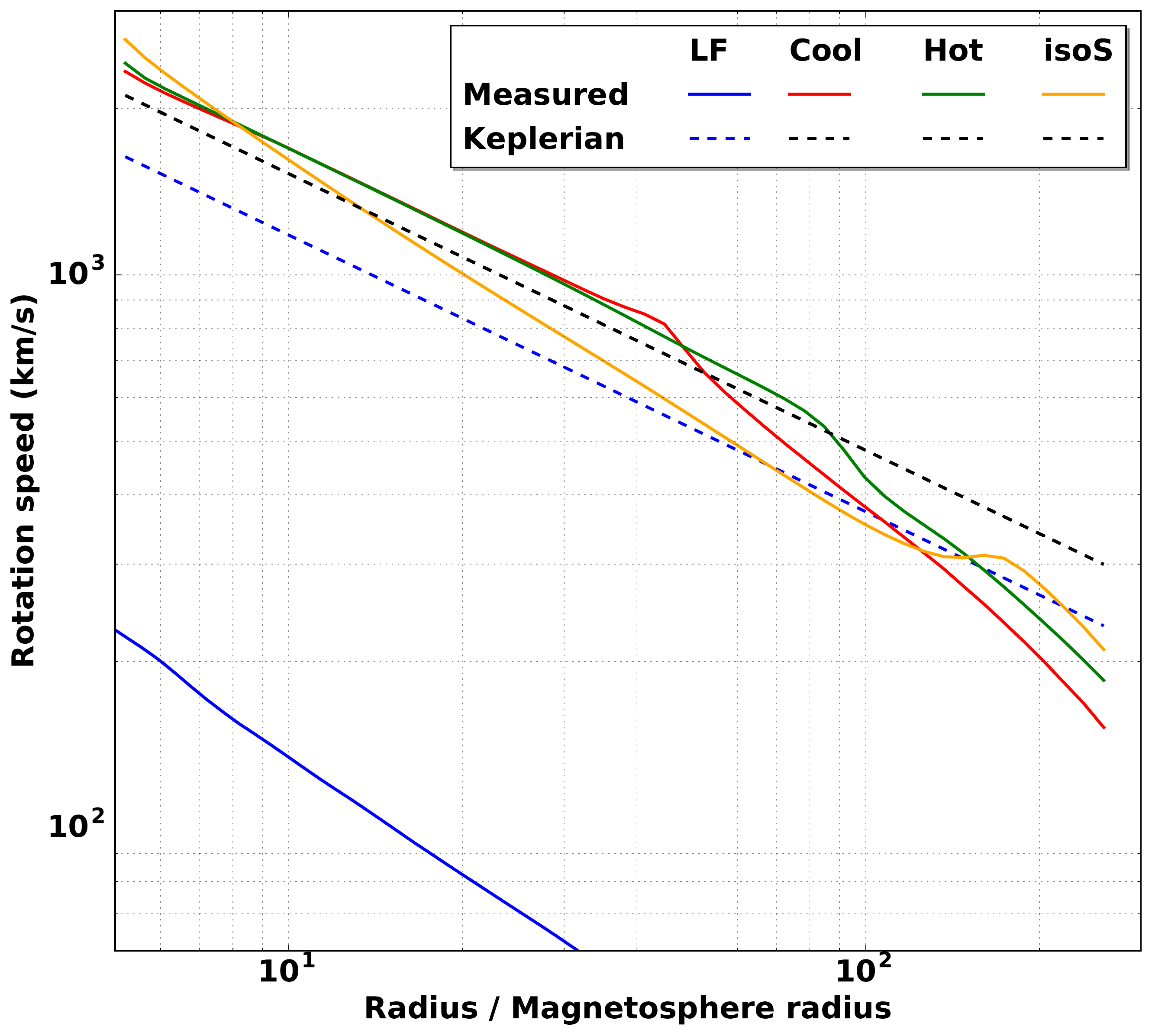}
\caption{Longitudinal velocity profiles in the orbital plane as a function of the distance to the accretor measured in units of the magnetosphere radius given in equation\,\eqref{eq:Rmag}. The velocities of the LF (blue) and HS configurations (red, green and yellow) have been measured once the numerically relaxed state has been reached and have been averaged over the longitudinal angles. The measured velocity profiles (solid lines) are compared to the Keplerian profiles expected for a thin disc around the light and the heavy \ns (dashed blue and dashed black respectively).}
\label{fig:vphi}
\end{figure} 

Let us now study its influence on the morphology of the flow in the LF and HS cases.

Whatever cooling prescription invoked, the LF setup never leads to the formation of a disc-like structure around the accretor. Instead, triggering the cooling for the LF flow leads to a serious recession of the front shock, down to the inner boundary of the simulation space, due to a drop of the pressure built-up downstream the shock. With the isentropic prescription, we ran a simulation with an inner boundary 5 times smaller to make sure that the size of the inner boundary was not impacting the morphology of the flow, and the result remained unchanged. Since the magnetic field is believed to play a role so close from the accretor and our simulations are only HD, we are not able to make any statement on the following accretion of the flow in the LF case, except that it is bound to not proceed via a disc. From now on, "LF" refers to the LF configuration coupled with the isentropic cooling prescription.

On the contrary, in the HS configuration, the front shock holds and a permanent disc-like structure forms within the shocked region, for all cooling prescription used (see Figure\,\ref{fig:cooled_ones}). In the isentropic case, the hull of the shock remains essentially unchanged, including the density bulge, since the cooling is only triggered in the innermost region, where the temperature of the flow goes beyond $T_0\sim 1$MK. There, we do observe the formation of a flattened persistent structure, partly supported by the centrifugal force (see Figure\,\ref{fig:disc}). As indicated by the relatively large thickness aspect ratio of the disc ($\sim$ 0.5), the pressure still plays an important role in sustaining the structure. Similarly, this disc-like structure appears in the two isothermal cases, with a thinner disc for a lower temperature. From now on, "cool", "hot" and "isoS" refer to the HS configuration coupled with the corresponding cooling prescription.

Another way to appreciate whether the flow is centrifugally supported and up to which radius is to plot the longitudinal velocity profile in the orbital plane. In Figure\,\ref{fig:vphi}, it is clear that the LF case displays a flow speed (solid blue line) very much below the Keplerian expectation (dashed blue), while the three HS cases show a much better matching between the measured (red, green and orange solid lines) and the Keplerian (black dashed) velocity profiles. Within the disc, the velocity profile is a power-law (the profiles are straight lines) but once we reach the outer extent of the disc, a sudden change in slope happens. A warmer disc corresponds to a larger extent of the disc. While the two isothermal cases decreases in $1/\sqrt{r}$ (in agreement with a constant temperature and a power-law density profile), we notice that they are both slightly offset from the Keplerian profile. They display rotational speed approximately 15\% above the Keplerian speed within the disc, which might be due to numerical effects at the inner border or representative of a not fully steady numerical state (see section\,\ref{sec:mdot_ldot}). In the isentropic case, we find a velocity profile decreasing faster than $1/\sqrt{r}$ which indicates an increasing importance of the thermal pressure in the equilibrium of this thicker disc.

Without cooling, we have seen that no disc formation is possible, independently of the net angular momentum carried by the accreted flow. In the context of disc formation during common envelope phase, \cite{MacLeod2015,MacLeod2017,Murguia-Berthier2017} performed similar simulations of asymmetric BHL accretion using a single polytropic relation everywhere in the domain. They concluded that disc formation was possible provided the flow was compressible enough, the compressibility being provided either by effective cooling or the capacity to convert a fraction of the energy released by compression to further ionize the stellar outer layers. Our results also agree with similar simulations of mass transfer in the context of binaries where the donor star is on the asymptotic giant branch \citep{HuarteEspinosa:2012wq,Saladino2018}. More generally, without energy loss, a flow with a given angular momentum can not circularize. By analogy with a test-mass, it would keep orbiting on the highly eccentric orbit the initial conditions imprinted. The shock mediates this analogy by adding entropy to the flow, but internal energy needs to be radiated away to lead to the formation of a centrifugally supported structure.

%
%

\section{Discussion and observational consequences}
\label{sec:obs_cons}

\subsection{Accretion rates}
\label{sec:mdot_ldot}

\subsubsection{Mass accretion rate}
\label{sec:mdot}

\begin{figure}
\centering
\includegraphics[width=0.95\columnwidth]{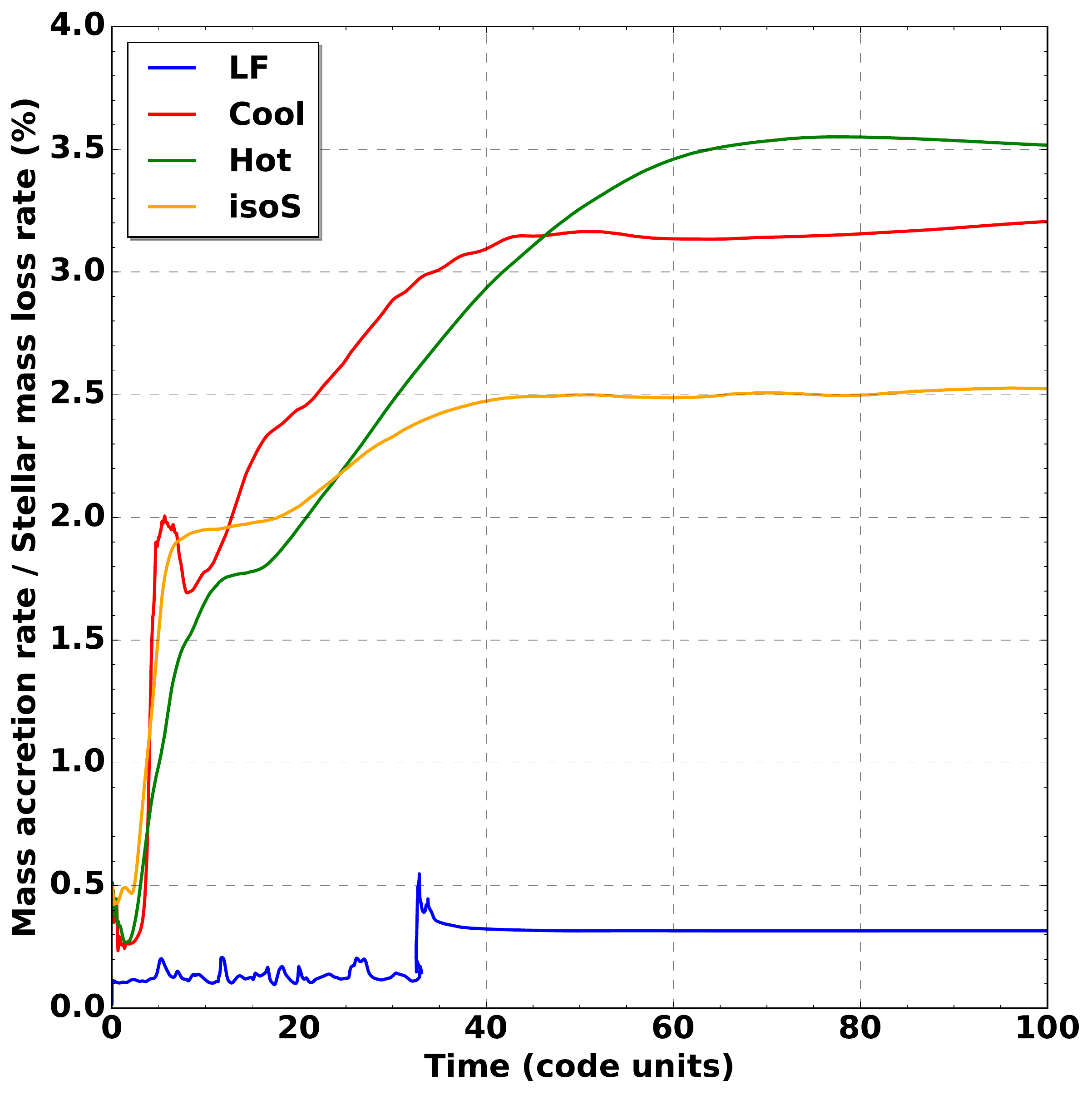}
\caption{Mass accretion rates at the inner border of the simulation space (at $\sim$5R$_{\text{mag}}$) as a function of time. In the case of the HS configuration, the mass accretion rate is significantly larger than in the LF case, whatever the cooling prescription.}
\label{fig:mdot_time}
\end{figure}

After at most a few crossing times, the total mass and angular momentum within the simulation space reach a plateau and the mass and angular momentum accretion rates do not vary by more than a few percents. The levels we observe depend significantly on the efficiency of the wind line-driven acceleration and on the mass of the accretor (LF and HS configurations). In Figure\,\ref{fig:mdot_time}, we represented the mass accretion rate at the inner boundary of the simulation space as a function of time. Comparing the HS and LF cases enables us to underline the dramatic increase in the mass accretion rate when the wind speed becomes similar to the orbital speed. While the mass accretion rate at the Roche lobe was approximately 4 times larger in the HS case (see section\,\ref{sec:orb_inhomo}), it is an order of magnitude higher in the HS case once it reaches the \ns magnetosphere. The mass accretion rates displayed in Figure\,\ref{fig:mdot_time} are steady and, within the shocked region, fairly independent of the radius at which they are measured, in agreement with the conservation of mass.

Currently, accretion of matter within the inner border is enabled only by the evacuation of angular momentum through spiral shocks (visible in Figure\,\ref{fig:disc}). However, in the absence of a proper treatment of the effective viscosity, statements on the absolute values of the mass accretion rate witnessed in these simulations should be taken with some caution. That being said, the values we observe, of the order of a few $\dot{M}\sim$10$^{-8}$M$_{\odot}\cdot$year$^{-1}$ would correspond to X-ray accretion luminosities of the order of $L_{\text{acc}}\sim\zeta\cdot3\cdot10^{39}$erg$\cdot$s$^{-1}$, where $\zeta$ encapsulates the information on the efficiency of the conversion process from kinetic energy to X-ray emission. Provided this coefficient reaches its maximum value given by the compactness parameter of the accretor (10 to 30\% for a \ns), the X-ray luminosity would be an order of magnitude higher than the Eddington luminosity L$_{\text{Edd}}$ of a \ns. In this case, accounting for the influence of the radiative force on the dynamics of the flow would be required to make reliable conclusions. Interestingly enough, it means that wind-RLOF, because it combines both the mass transfer efficiency of RLOF (usually associated to LMXB) and the large amount of available matter provided by the mass loss mechanism of a massive star (associated to HMXB), could lead to super-Eddington accretion. In particular, we notice that some Ultra Luminous X-ray sources (ULX) have been shown to be super-Eddington accreting \ns \citep{Bachetti2014,Furst2016,Israel2017}. Supergiant or Wolf-Rayet donor stars such as the ones identified in some systems could provide material at a rate sufficient to reproduce this X-ray luminosity \citep[][submitted]{ElMellah2018a}. 

Regarding the current systems of interest in this paper, classic \sgx and Vela X-1 in particular, it would be inconsistent with the observed X-ray luminosity ranging from 10$^{35}$erg$\cdot$s$^{-1}$ to a few 10$^{37}$erg$\cdot$s$^{-1}$ in Vela X-1 \citep{Furst2010}. In the case of an accreting black hole, very low values of $\zeta$ can be reached for radiatively inefficient accretion flow, either because the flow has no time to cool or because it is optically thick enough to drag the radiation in its fall towards the event horizon \citep{Narayan1998}. But for a \ns, the kinetic energy must be released before or at the impact with the \ns surface \citep{Medvedev2000}. Very faint accreting \ns have been identified in \lmxb \citep{ArmasPadilla2013}. In spite of a \rlof mass transfer susceptible to provide a large fraction of the Eddington mass accretion rate, these systems maintain low X-ray luminosities of 5$\cdot$10$^{34}$ to 5$\cdot$10$^{36}$erg$\cdot$s$^{-1}$ (\ie 10$^{-4}$ to 10$^{-2}$L$_{\text{Edd}}$) for years, much below the levels generally observed in LMXB. \cite{Degenaar2017} reported about observational signatures of an outflow in one of these sources, IGR J17062–6143, and suggested two mechanisms to account for the low luminosity of the accreting \ns. The first possibility could be a truncated disc with a radiatively-inefficient accretion flow in the inner parts \citep[see \eg the adiabatic inflow-outflow solution derived by][]{Blandford1999}, possibly associated with an outflow. An alternative scenario is a propeller-driven outflow. In the regions where we monitored the flow, the magnetic field carried by the flow has little influence on the motion of the gas. However, when the flow gets close enough from the accretor, it gets highly ionized by the X-ray emission and encounters the intense dipolar magnetic field of the \ns. From this point, the magnetic field takes over and controls the dynamics : in \sgx, any putative disc-like structure would eventually be truncated way before the \ns surface \citep{Ghosh1978}. In the propeller regime, part of the matter falling onto the magnetosphere might eventually be repelled, leading to a much lower effective mass accretion rate than the amount inflowing at a few magnetosphere radii \citep{Illarionov1975,Bozzo2008}.

%
%


\subsubsection{Angular momentum accretion rate}
\label{sec:ldot}

When a disc forms, the angular momentum accretion rate at a few R$_{\text{mag}}$ can in no case be indicative of the torque applied to the accretor since it is precisely thanks to the evacuation of a significant amount of angular momentum that accretion takes place. However, it provides an upper limit which can be computed to monitor the consistency of these simulations. In addition, the characteristics of the exchange of angular momentum between the flow and the \ns depends on the extent of the corotation radius with respect to the magnetosphere radius \citep{Ghosh1979}. A naive computation of the characteristic spin-up time $\tau_{\text{L}}$ of a \ns of mass M$_{\bullet}=2$M$_{\odot}$, of radius R$_{\bullet}=10$km and of spin period $P_{\bullet}=2\pi/\omega=283$s, representative of the one hosted in Vela X-1, would give, with the angular momentum accretion rate $\dot{L}$ we measure at the inner edge of the HS simulations :
\begin{equation}
\tau_{\text{L}}=\frac{L_{\bullet}}{\dot{L}}=\frac{M_{\bullet}R_{\bullet}^2\omega_{\bullet}^2}{\dot{L}}\sim 10\text{kyrs}
\end{equation}
which is one to three orders of magnitude smaller than the net spin-up/down time observed reported in the literature \citep[see \eg][]{Ziolkowski1985}. It shows that the angular momentum flowing through the inner boundary of our simulation space is too large and that only the fraction which is not evacuated during the accretion process will actually reach the \ns and play a role in spinning it up/down.

\subsection{Disc extension and viscous lag : the case of Cygnus X-1}
\label{sec:rout}

Even in systems which are known to harbor an accretion disc, we have often little insight on the disc outer radius. Indeed, most of the light comes from the innermost region, which are also, for a disc extending all the way down to the compact accretor, the ones emitting in X-rays, in a waveband well separated from the stellar black body emission. In \rlof systems, the monitoring of the hot spot can help to locate this outer ring, provided the disc is not tilted, and theoretical limits exist on the maximum extension of the disc \citep[$\sim$70\% of the Roche lobe radius of the accretor,][]{Paczynski1977}. 

Recently, \cite{Taam2018} invoked a hot, low angular momentum accretion flow to explain the absence of visible hysteresis in the hardness-intensity cycle of Cygnus X-1, where the contrast is low between the brightest and dimmest X-ray emission compared to BH in LMXB \citep{Grinberg:2014ux}. Indeed, in Cygnus X-1, mass transfer most likely proceeds through wind accretion. The present study, although focused on the case of a \ns accretor, indicates that, if the wind speed is not negligible compared to the orbital speed, the disc should be smaller than what a \rlof process would produce in a classic BH-LMXB. In Cygnus X-1, the donor star is of similar mass as in Vela X-1 but with a radius smaller by almost a factor of 2 \citep{Orosz2011}, which would lead, for a similar Eddington parameter, to a larger wind terminal speed if it scales as the effective escape speed at the stellar photosphere (as discussed in section\,\ref{sec:orb_model}). But this increase must be compared to the orbital speed in Cygnus X-1 ($\sim$400km$\cdot$s$^{-1}$) which is also larger than in Vela X-1, and to the much more important mass of the accretor relative to the donor star. All in all, the extent of the wind-captured disc formed in Cygnus X-1 might be larger than observed in the HS configuration, more suitable for Vela X-1, but much smaller than a \rlof formed disc. \cite{Smith2001} already pinpointed that the delay between the hard to soft and soft to hard transitions might be representative of the different propagation times in a two components accretion flow : while the corona reacts to a perturbation of a free-fall timescale, the disc, limited by the viscous timescale, would always lag behind. Because the viscous timescale depends on the outer extent of the disc, a lower delay in the hardness-intensity cycle of Cygnus X-1 compared to BH-LMXB would indicate a smaller disc extent, consistent with the present study of the properties of wind-captured discs.

\section{Conclusion}
\label{sec:conc}

In this paper, we tried to evaluate the possibility to obtain a wind-captured disc in \sgx and in particular in the classic \ns -hosting \sgx Vela X-1. We connected the orbital scale, at which the wind develops, to the one of the accretion radius, at which the flow is significantly beamed by the gravitational field of the compact object and where HD shocks form, all the way down to the outer edge of the \ns magnetosphere. With these simulations, we consistently cover the development of the wind as it is launched, compute its deviation at the orbital scale and evaluate the fraction eventually reaching the \ns magnetosphere. We showed that the wind dramatically departs from a radial outflow when the wind speed is as low as the orbital speed. We also capture the adiabatic bow shock which forms ahead of the accretor and characterize its highly asymmetric shape for a slow wind. Provided cooling is triggered in the shocked region, the accreted flow formed out of a slow wind circularizes at a few 10 times the \ns magnetosphere radius. The obtained disc-like structure is essentially maintained by the centrifugal force and displays a quasi-Keplerian rotation profile.

Currently, we face a lack of conclusive evidence of the presence of a permanent disc in \sgx hosting \ns \citep{Bozzo2008,Shakura2012,Romano2015,Hu2017}. Yet, because of the truncation of the disc by the \ns magnetosphere \citep{Ghosh1978}, we do not expect from it an emission as intense and as high energy as for a disc extending deeply into the gravitational potential of the compact accretor. In UV waveband, the emission from a putative disc would be dominated by the flux from the O/B Sg star and in the X-rays, accretion columns at the \ns poles would be much brighter than the truncated disc. Alternatively, if the wind actually reaches speeds larger than the orbital speed upstream the accretor, a wind-captured disc would be ruled out by the present study. \cite{Taani2018} recently identified possible hints of accretion via a disc in 2 \sgx, OAO 1657-415 and LMC X-4. The latter has already been pinpointed in the literature as a possible disc-fed \sgx : based on their position in the lower left end of the Corbet diagram \citep{Corbet1984} and on their high X-ray luminosity, a disc is believed to form around the compact object in LMC X-4 but also in Cen X-3 or SMC X-1 \citep{Falanga2015}. However, the large mass ratios ($>10$) in these systems suggest that a stable pure RLOF is unlikely, a inconsistency which could be solved by the present mass transfer mechanism : wind-RLOF remains stable for large mass ratios while still leading, for realistically slow winds, to the formation of a wind-captured disc around the accretor.

To summarize, in Vela X-1, we face two possible wind accretion scenarios :
\begin{itemize}
\item either the line-driven wind acceleration is not efficient enough to provide the flow with a velocity larger than the orbital speed when it enters the Roche lobe of the accretor. In this case, the orbital effects control the dynamics of the wind which is seriously beamed, with a fraction of the stellar mass loss rate entering the Roche lobe of the accretor larger than 10\%. The flow quickly acquires angular momentum and within the shocked region, a disc-like structure is formed, with thicker discs for less efficient cooling. Because Vela X-1 is an eclipsing system and since we expect the wind-captured disc to form within the orbital plane, the thickness of the disc might significantly alter the absorbing column density : in the case of an inefficient cooling, the disc would be thick enough to intercept the line-of-sight and contribute to the observed absorbing column density. Also, \cite{Foulkes:2010wa} showed that the precession of a disc wrapped by a radiative pressure driven instability \citep{Petterson1977a,Petterson1977} could be responsible for the off-states observed in some systems, although mechanisms not relying on the presence of a disc have been proposed for Vela X-1 \citep[see \eg][]{Manousakis2015c}.
\item or the line-driven wind acceleration is more efficient than the one computed for a 1D steady atmosphere in \cite{Sander2017}, for instance because of a significant influence of the departure from a purely radial wind or a different chemical composition of the donor star than assumed. For instance, among the most important driving elements of the donor star in Vela X-1, the abundances of N and Si are determined from spectral lines, but the abundances of S and Fe are just assumed to be solar \citep{Gimenez-Garcia2016} : since the Fe forest is well reproduced though, the abundances are thought to be accurate within a factor of at most 2. In the case of a more efficient wind acceleration, the LF configuration, with a wind speed slightly larger than the orbital speed, would be more representative of the wind accretion process at stake in Vela X-1 and no disc structure would form. The mass accretion rate, of the order of a few 0.1\% of the stellar mass loss rate, would be consistent with the observed mean X-ray luminosity for typical values of $\zeta$ (\ie $\sim$ 5 to 10\%).
\end{itemize}

The present study addressed the permanent behavior of the flow, but winds of massive stars are unstable and form internal shocks which lead to overdense clumps of matter \citep{Owocki1984,Sundqvist2017}. The impact of these clumps on the time variability of the mass accretion rate has been investigated in \cite{ElMellah} but for a wind fast enough to not be bent by the orbital effects was considered. The present work shows that, provided the wind speed is similar to the orbital speed, the orbital bending and the associated net angular momentum can not be neglected. Because the clumps carry themselves a local angular momentum, \cite{ElMellah} showed that clumpy wind accretion was less efficient than its homogeneous counterpart : including the micro-structure of the wind might thus help to decrease the mass accretion rate at a few \ns magnetosphere radii, but the influence of the clumps on the properties of the wind-captured disc (\eg its permanence) remains to be studied. With a proper treatment of the effective viscosity in the disc and of the coupling between the disc and the magnetosphere, we could produce a physically-motivated synthetic curve of the torque applied to the \ns as a function of time, opening the door to a consistent way to address the question of the \ns spinning-up and down.

\begin{acknowledgements}
IEM wishes to thank Antonios Manousakis for fruitful discussions about the underlying computational aspects and the members of the X-wind collaboration for the observational consequences of the present analysis. IEM also thanks Wenbin Liu for the insightful discussions on the accretion of angular momentum. IEM has received funding from the Research Foundation Flanders (FWO) and the European Union's Horizon 2020 research and innovation program under the Marie Sk\l odowska-Curie grant agreement No 665501. A.A.C.S. would like to thank STFC for funding under grant number ST/R000565/1. IEM and JOS are grateful for the hospitality of the International Space Science Institute (ISSI), Bern, Switzerland which sponsored a team meeting initiating a tighter collaboration between massive stars wind and X-ray binaries communities. The simulations were conducted on the Tier-1 VSC (Flemish Supercomputer Center funded by Hercules foundation and Flemish government).
\end{acknowledgements}


\bibliographystyle{aa} 
\begin{tiny}
\bibliography{article_sheared_wind_no_url}

\begin{thebibliography}{82}
\expandafter\ifx\csname natexlab\endcsname\relax\def\natexlab#1{#1}\fi

\bibitem[{Abbott {et~al.}(2016)Abbott, Abbott, Abbott, \& Al}]{Abbott2016a}
Abbott, B.~P., Abbott, R., Abbott, T.~D., \& Al, E. 2016, Phys. Rev. Lett.,
  116, 241103

\bibitem[{{Armas Padilla} {et~al.}(2013){Armas Padilla}, Degenaar, \&
  Wijnands}]{ArmasPadilla2013}
{Armas Padilla}, M., Degenaar, N., \& Wijnands, R. 2013, Mon. Not. R. Astron.
  Soc., 434, 1586

\bibitem[{Bachetti {et~al.}(2014)Bachetti, Harrison, Walton, Grefenstette,
  Chakrabarty, F{\"{u}}rst, Barret, Beloborodov, Boggs, Christensen, Craig,
  Fabian, Hailey, Hornschemeier, Kaspi, Kulkarni, MacCarone, Miller, Rana,
  Stern, Tendulkar, Tomsick, Webb, \& Zhang}]{Bachetti2014}
Bachetti, M., Harrison, F.~A., Walton, D.~J., {et~al.} 2014, Nature, 514, 202

\bibitem[{Blandford \& Begelman(1999)}]{Blandford1999}
Blandford, R.~D. \& Begelman, M.~C. 1999, Mon. Not. R. Astron. Soc., 303, 1

\bibitem[{Blondin {et~al.}(1990)Blondin, Kallman, Fryxell, \&
  Taam}]{Blondin1990a}
Blondin, J.~M., Kallman, T.~R., Fryxell, B.~A., \& Taam, R.~E. 1990, Astrophys.
  J., 356, 591

\bibitem[{Blondin \& Raymer(2012)}]{Blondin:2012vf}
Blondin, J.~M. \& Raymer, E. 2012, Astrophys. J., 752, 30

\bibitem[{Bozzo {et~al.}(2008)Bozzo, Falanga, \& Stella}]{Bozzo2008}
Bozzo, E., Falanga, M., \& Stella, L. 2008, Astrophys. J., 683, 1031

\bibitem[{Castor {et~al.}(1975)Castor, Abbott, \& Klein}]{Castor1975}
Castor, J.~I., Abbott, D.~C., \& Klein, R.~I. 1975, Astrophys. J., 195, 157

\bibitem[{Christians(2012)}]{Christians2012}
Christians, J. 2012, Int. J. Mech. Eng. Educ., 40, 53

\bibitem[{Corbet(1984)}]{Corbet1984}
Corbet, R. H.~D. 1984, Astron. Astrophys. (ISSN 0004-6361), vol. 141, no. 1,
  Dec. 1984, p. 91-93., 141, 91

\bibitem[{Degenaar {et~al.}(2017)Degenaar, Pinto, Miller, Wijnands, Altamirano,
  Paerels, Fabian, \& Chakrabarty}]{Degenaar2017}
Degenaar, N., Pinto, C., Miller, J.~M., {et~al.} 2017, Mon. Not. R. Astron.
  Soc., 464, 398

\bibitem[{Duch{\^{e}}ne \& Kraus(2013)}]{Duchene2013}
Duch{\^{e}}ne, G. \& Kraus, A. 2013, Annu. Rev. Astron. Astrophys., 51, 269

\bibitem[{Edgar(2004)}]{Edgar:2004ip}
Edgar, R.~G. 2004, New Astron. Rev., 48, 843

\bibitem[{{El Mellah} \& Casse(2015)}]{ElMellah2015}
{El Mellah}, I. \& Casse, F. 2015, Mon. Not. R. Astron. Soc., 454, 2657

\bibitem[{{El Mellah} \& Casse(2016)}]{ElMellah2016a}
{El Mellah}, I. \& Casse, F. 2016, Mon. Not. R. Astron. Soc., 467, 2585

\bibitem[{{El Mellah} {et~al.}(2017){El Mellah}, Sundqvist, \&
  Keppens}]{ElMellah}
{El Mellah}, I., Sundqvist, J.~O., \& Keppens, R. 2017, Mon. Not. R. Astron.
  Soc., 475, 3240

\bibitem[{{El Mellah} {et~al.}(2018){El Mellah}, Sundqvist, \&
  Keppens}]{ElMellah2018a}
{El Mellah}, I., Sundqvist, J.~O., \& Keppens, R. 2018, (submitted)

\bibitem[{Falanga {et~al.}(2015)Falanga, Bozzo, Lutovinov, Bonnet-Bidaud,
  Fetisova, \& Puls}]{Falanga2015}
Falanga, M., Bozzo, E., Lutovinov, A., {et~al.} 2015, Astron. Astrophys., 577,
  A130

\bibitem[{Foglizzo {et~al.}(2005)Foglizzo, Galletti, \& Ruffert}]{Foglizzo2005}
Foglizzo, T., Galletti, P., \& Ruffert, M. 2005, Astron. Astrophys., 2201, 15

\bibitem[{Foglizzo \& Ruffert(1996)}]{Foglizzo1996}
Foglizzo, T. \& Ruffert, M. 1996, Astron. Astrophys., 361, 22

\bibitem[{Forman {et~al.}(1973)Forman, Jones, Tananbaum, Gursky, Kellogg, \&
  Giacconi}]{Forman1973}
Forman, W., Jones, C., Tananbaum, H., {et~al.} 1973, Astrophys. J., 182, L103

\bibitem[{Foulkes {et~al.}(2010)Foulkes, Haswell, \& Murray}]{Foulkes:2010wa}
Foulkes, S.~B., Haswell, C.~A., \& Murray, J.~R. 2010, Mon. Not. R. Astron.
  Soc., 401, 1275

\bibitem[{Frank {et~al.}(1986)Frank, King, \& Raine}]{Frank2002}
Frank, J., King, A., \& Raine, D.~J. 1986, Phys. Today, 39, 124

\bibitem[{F{\"{u}}rst {et~al.}(2018)F{\"{u}}rst, Kretschmar, Grinberg,
  Pottschmidt, Wilms, K{\"{u}}hnel, {El Mellah}, \&
  Mart{\'{i}}nez-N{\'{u}}{\~{n}}ez}]{Furst2018}
F{\"{u}}rst, F., Kretschmar, P., Grinberg, V., {et~al.} 2018

\bibitem[{F{\"{u}}rst {et~al.}(2010)F{\"{u}}rst, Kreykenbohm, Pottschmidt,
  Wilms, Hanke, Rothschild, Kretschmar, Schulz, Huenemoerder, Klochkov, \&
  Staubert}]{Furst2010}
F{\"{u}}rst, F., Kreykenbohm, I., Pottschmidt, K., {et~al.} 2010, Astron.
  Astrophys., 519, A37

\bibitem[{F{\"{u}}rst {et~al.}(2014)F{\"{u}}rst, Pottschmidt, Wilms, Tomsick,
  Bachetti, Boggs, Christensen, Craig, Grefenstette, Hailey, Harrison, Madsen,
  Miller, Stern, Walton, \& Zhang}]{Furst2014}
F{\"{u}}rst, F., Pottschmidt, K., Wilms, J., {et~al.} 2014, Astrophys. J., 780
  [\eprint[arXiv]{1311.5514}]

\bibitem[{F{\"{u}}rst {et~al.}(2016)F{\"{u}}rst, Walton, Harrison, Stern,
  Barret, Brightman, Fabian, Grefenstette, Madsen, Middleton, Miller,
  Pottschmidt, Ptak, Rana, \& Webb}]{Furst2016}
F{\"{u}}rst, F., Walton, D.~J., Harrison, F.~A., {et~al.} 2016, Astrophys. J.,
  831, L14

\bibitem[{Ghosh \& Lamb(1978)}]{Ghosh1978}
Ghosh, P. \& Lamb, F.~K. 1978, Astrophys. J., 223, L83

\bibitem[{Ghosh \& Lamb(1979)}]{Ghosh1979}
Ghosh, P. \& Lamb, F.~K. 1979, Astrophys. J., 234, 296

\bibitem[{Gimenez-Garcia {et~al.}(2016)Gimenez-Garcia, Shenar, Torrejon,
  Oskinova, Martinez-Nunez, Hamann, Rodes-Roca, Gonzalez-Galan,
  Alonso-Santiago, Gonzalez-Fernandez, Bernabeu, \&
  Sander}]{Gimenez-Garcia2016}
Gimenez-Garcia, A., Shenar, T., Torrejon, J.~M., {et~al.} 2016, Astron.
  Astrophys., 591, 25

\bibitem[{Gr{\"{a}}fener {et~al.}(2002)Gr{\"{a}}fener, Koesterke, \&
  Hamann}]{Grafener2002}
Gr{\"{a}}fener, G., Koesterke, L., \& Hamann, W.-R. 2002, Astron. Astrophys.,
  387, 244

\bibitem[{Grinberg {et~al.}(2017)Grinberg, Hell, {El Mellah}, Neilsen, Sander,
  Leutenegger, F{\"{u}}rst, Huenemoerder, Kretschmar, K{\"{u}}hnel,
  Mart{\'{i}}nez-N{\'{u}}{\~{n}}ez, Niu, Pottschmidt, Schulz, Wilms, \&
  Nowak}]{Grinberg2017}
Grinberg, V., Hell, N., {El Mellah}, I., {et~al.} 2017, Astron. Astrophys. Vol.
  608, id.A143, 18 pp., 608 [\eprint[arXiv]{1711.06743}]

\bibitem[{Grinberg {et~al.}(2014)Grinberg, Pottschmidt, B{\"{o}}ck, Schmid,
  Nowak, Uttley, Tomsick, Rodriguez, Hell, Markowitz, Bodaghee, Bel,
  Rothschild, \& Wilms}]{Grinberg:2014ux}
Grinberg, V., Pottschmidt, K., B{\"{o}}ck, M., {et~al.} 2014, Astron.
  Astrophys. Vol. 565, id.A1, 19 pp., 565 [\eprint[arXiv]{1402.4485}]

\bibitem[{Hamann \& Koesterke(1998)}]{Hamann1998}
Hamann, W.-R. \& Koesterke, L. 1998, Astron. Astrophys, 335, 1003

\bibitem[{Hatchett \& McCray(1977)}]{Hatchett1977}
Hatchett, S. \& McCray, R. 1977, Astrophys. J., 211, 552

\bibitem[{Hiltner {et~al.}(1972)Hiltner, Werner, \& Osmer}]{Hiltner1972}
Hiltner, W.~A., Werner, J., \& Osmer, P. 1972, Astrophys. J., 175, L19

\bibitem[{Ho \& Arons(1987)}]{Ho1987}
Ho, C. \& Arons, J. 1987, Astrophys. J., 316, 283

\bibitem[{Horedt(2000)}]{Horedt2000}
Horedt, G.~P. 2000, Astrophys. J., 541, 821

\bibitem[{Hu {et~al.}(2017)Hu, Chou, Ng, Lin, \& Yen}]{Hu2017}
Hu, C.-P., Chou, Y., Ng, C.-Y., Lin, L. C.-C., \& Yen, D. C.-C. 2017,
  Astrophys. J., 844, 16

\bibitem[{Huarte-Espinosa {et~al.}(2012)Huarte-Espinosa, Carroll-Nellenback,
  Nordhaus, Frank, \& Blackman}]{HuarteEspinosa:2012wq}
Huarte-Espinosa, M., Carroll-Nellenback, J., Nordhaus, J., Frank, A., \&
  Blackman, E.~G. 2012, Mon. Not. R. Astron. Soc., 433, 295

\bibitem[{Illarionov \& Sunyaev(1975)}]{Illarionov1975}
Illarionov, A.~F. \& Sunyaev, R.~A. 1975, Astron. Astrophys., 39

\bibitem[{Israel {et~al.}(2017)Israel, Belfiore, Stella, Esposito, Casella, {De
  Luca}, Marelli, Papitto, Perri, Puccetti, Castillo, Salvetti, Tiengo,
  Zampieri, D'Agostino, Greiner, Haberl, Novara, Salvaterra, Turolla, Watson,
  Wilms, Wolter, {Rodr{\'{i}}guez Castillo}, Salvetti, Tiengo, Zampieri,
  D'Agostino, Greiner, Haberl, Novara, Salvaterra, Turolla, Watson, Wilms, \&
  Wolter}]{Israel2017}
Israel, G.~L., Belfiore, A., Stella, L., {et~al.} 2017, Science (80-. )., 355,
  817

\bibitem[{Karino(2014)}]{Karino2014}
Karino, S. 2014, Publ. Astron. Soc. Japan, 66, 2

\bibitem[{Lucy \& Solomon(1970)}]{Lucy1970}
Lucy, L.~B. \& Solomon, P.~M. 1970, Astrophys. J., 159, 879

\bibitem[{MacLeod {et~al.}(2017)MacLeod, Antoni, Murgia-Berthier, Macias, \&
  Ramirez-Ruiz}]{MacLeod2017}
MacLeod, M., Antoni, A., Murgia-Berthier, A., Macias, P., \& Ramirez-Ruiz, E.
  2017, Astrophys. J., 838, 56

\bibitem[{MacLeod \& Ramirez-Ruiz(2015)}]{MacLeod2015}
MacLeod, M. \& Ramirez-Ruiz, E. 2015, Astrophys. J., 803, 41

\bibitem[{Manousakis \& Walter(2015)}]{Manousakis2015c}
Manousakis, A. \& Walter, R. 2015, Astron. Astrophys., 58, A58

\bibitem[{Manousakis {et~al.}(2014)Manousakis, Walter, \&
  Blondin}]{Manousakis2014}
Manousakis, A., Walter, R., \& Blondin, J. 2014, EPJ Web Conf., 64, 02006

\bibitem[{Mart{\'{i}}nez-N{\'{u}}{\~{n}}ez
  {et~al.}(2017)Mart{\'{i}}nez-N{\'{u}}{\~{n}}ez, Kretschmar, Bozzo, Oskinova,
  Puls, Sidoli, Sundqvist, Blay, Falanga, F{\"{u}}rst,
  G{\'{i}}menez-Garc{\'{i}}a, Kreykenbohm, K{\"{u}}hnel, Sander,
  Torrej{\'{o}}n, \& Wilms}]{Martinez-Nunez2017}
Mart{\'{i}}nez-N{\'{u}}{\~{n}}ez, S., Kretschmar, P., Bozzo, E., {et~al.} 2017,
  Space Sci. Rev., 212, 59

\bibitem[{Medvedev \& Narayan(2000)}]{Medvedev2000}
Medvedev, M.~V. \& Narayan, R. 2000, Astrophys. Journal, Vol. 554, Issue 2, pp.
  1255-1267., 554, 1255

\bibitem[{Mohamed \& Podsiadlowski(2007)}]{Mohamed2007}
Mohamed, S. \& Podsiadlowski, P. 2007

\bibitem[{Molteni {et~al.}(1999)Molteni, T{\'{o}}th, \&
  Kuznetsov}]{Molteni1999}
Molteni, D., T{\'{o}}th, G., \& Kuznetsov, O. A.~O. 1999, Apj, 516, 411

\bibitem[{Murguia-Berthier {et~al.}(2017)Murguia-Berthier, MacLeod,
  Ramirez-Ruiz, Antoni, \& Macias}]{Murguia-Berthier2017}
Murguia-Berthier, A., MacLeod, M., Ramirez-Ruiz, E., Antoni, A., \& Macias, P.
  2017, Astrophys. J., 845, 173

\bibitem[{Narayan {et~al.}(1998)Narayan, Mahadevan, \& Quataert}]{Narayan1998}
Narayan, R., Mahadevan, R., \& Quataert, E. 1998, Theory Black Hole Accretion
  Disk. - Cambridge Univ. Press, 36

\bibitem[{Orosz {et~al.}(2011)Orosz, McClintock, Aufdenberg, Remillard, Reid,
  Narayan, \& Gou}]{Orosz2011}
Orosz, J.~A., McClintock, J.~E., Aufdenberg, J.~P., {et~al.} 2011, Astrophys.
  J., 742, 84

\bibitem[{Owocki \& Rybicki(1984)}]{Owocki1984}
Owocki, S.~P. \& Rybicki, G.~B. 1984, Astrophys. J., 284, 337

\bibitem[{Paczynski(1977)}]{Paczynski1977}
Paczynski, B. 1977, Astrophys. J., 216, 822

\bibitem[{Petterson \& A.(1977{\natexlab{a}})}]{Petterson1977a}
Petterson, J.~A. \& A., J. 1977{\natexlab{a}}, Astrophys. J., 214, 550

\bibitem[{Petterson \& A.(1977{\natexlab{b}})}]{Petterson1977}
Petterson, J.~A. \& A., J. 1977{\natexlab{b}}, Astrophys. J., 216, 827

\bibitem[{Quaintrell {et~al.}(2003)Quaintrell, Norton, Ash, Roche, Willems,
  Bedding, Baldry, \& Fender}]{Quaintrell2003a}
Quaintrell, H., Norton, A.~J., Ash, T. D.~C., {et~al.} 2003, Astron.
  Astrophys., 401, 313

\bibitem[{Rawls {et~al.}(2011)Rawls, Orosz, McClintock, Torres, Bailyn, \&
  Buxton}]{Rawls2011}
Rawls, M.~L., Orosz, J.~A., McClintock, J.~E., {et~al.} 2011, Astrophys.
  Journal, Vol. 730, Issue 1, Artic. id. 25, 11 pp. (2011)., 730
  [\eprint[arXiv]{1101.2465}]

\bibitem[{Romano {et~al.}(2015)Romano, Bozzo, Mangano, Esposito, Israel,
  Tiengo, Campana, Ducci, Ferrigno, \& Kennea}]{Romano2015}
Romano, P., Bozzo, E., Mangano, V., {et~al.} 2015, Astron. Astrophys., 576, 5

\bibitem[{Saladino {et~al.}(2018)Saladino, Pols, van~der Helm, Pelupessy, \&
  Zwart}]{Saladino2018}
Saladino, M.~I., Pols, O.~R., van~der Helm, E., Pelupessy, I., \& Zwart, S.~P.
  2018, eprint arXiv:1805.03208 [\eprint[arXiv]{1805.03208}]

\bibitem[{Sana {et~al.}(2012)Sana, de~Mink, de~Koter, Langer, Evans, Gieles,
  Gosset, Izzard, {Le Bouquin}, Schneider, Bouquin, \& Schneider}]{Sana2012}
Sana, H., de~Mink, S.~E., de~Koter, A., {et~al.} 2012, Sci. Vol. 337, Issue
  6093, pp. 444- (2012)., 337, 444

\bibitem[{Sander {et~al.}(2017{\natexlab{a}})Sander, F{\"{u}}rst, Kretschmar,
  Oskinova, Todt, Hainich, Shenar, \& Hamann}]{Sander2017}
Sander, A. A.~C., F{\"{u}}rst, F., Kretschmar, P., {et~al.} 2017{\natexlab{a}},
  Astron. Astrophys. Vol. 610, id.A60, 19 pp., 610, A60

\bibitem[{Sander {et~al.}(2017{\natexlab{b}})Sander, Hamann, Todt, Hainich, \&
  Shenar}]{Sander2017b}
Sander, A. A.~C., Hamann, W.-R. W.-R., Todt, H., Hainich, R., \& Shenar, T.
  2017{\natexlab{b}}, Astron. Astrophys. Vol. 603, id.A86, 14 pp., 603
  [\eprint[arXiv]{1704.08698}]

\bibitem[{Schure {et~al.}(2009)Schure, Kosenko, Kaastra, Keppens, \&
  Vink}]{Schure2009}
Schure, K.~M., Kosenko, D., Kaastra, J.~S., Keppens, R., \& Vink, J. 2009,
  Astron. Astrophys. Vol. 508, Issue 2, 2009, pp.751-757, 757, 751

\bibitem[{Shakura {et~al.}(2012)Shakura, Postnov, Kochetkova, \&
  Hjalmarsdotter}]{Shakura2012}
Shakura, N., Postnov, K., Kochetkova, A., \& Hjalmarsdotter, L. 2012, Mon. Not.
  R. Astron. Soc., 420, 216

\bibitem[{Shakura {et~al.}(2013)Shakura, Postnov, Kochetkova, \&
  Hjalmarsdotter}]{Shakura2013b}
Shakura, N.~I., Postnov, K.~A., Kochetkova, A.~Y., \& Hjalmarsdotter, L. 2013,
  Physics-Uspekhi, 56, 321

\bibitem[{Shapiro \& Lightman(1976)}]{Shapiro1976}
Shapiro, S.~L. \& Lightman, A.~P. 1976, Astrophys. J., 204, 555

\bibitem[{Smith {et~al.}(2001)Smith, Heindl, \& Swank}]{Smith2001}
Smith, D.~M., Heindl, W.~A., \& Swank, J.~H. 2001, Astrophys. Journal, Vol.
  569, Issue 1, pp. 362-380., 569, 362

\bibitem[{Sundqvist {et~al.}(2017)Sundqvist, Owocki, \& Puls}]{Sundqvist2017}
Sundqvist, J.~O., Owocki, S.~P., \& Puls, J. 2017, Astron. Astrophys. Vol. 611,
  id.A17, 10 pp., 611 [\eprint[arXiv]{1710.07780}]

\bibitem[{Taam {et~al.}(2018)Taam, Qiao, Liu, \& Meyer-Hofmeister}]{Taam2018}
Taam, R.~E., Qiao, E., Liu, B.~F., \& Meyer-Hofmeister, E. 2018, Astrophys.
  Journal, Vol. 860, Issue 2, Artic. id. 166, 5 pp. (2018)., 860, 1

\bibitem[{Taani {et~al.}(2018)Taani, Karino, Song, Al-Wardat, Khasawneh, \&
  Mardini}]{Taani2018}
Taani, A., Karino, S., Song, L., {et~al.} 2018, eprint arXiv:1809.01213
  [\eprint[arXiv]{1809.01213}]

\bibitem[{Toro {et~al.}(1994)Toro, Spruce, \& Speares}]{Toro1994}
Toro, E.~F., Spruce, M., \& Speares, W. 1994, Shock Waves, 4, 25

\bibitem[{van Marle \& Keppens(2011)}]{VanMarle2011}
van Marle, A.~J. \& Keppens, R. 2011, Comput. Fluids, 42, 44

\bibitem[{Vink {et~al.}(2001)Vink, de~Koter, \& Lamers}]{Vink2001}
Vink, J.~S., de~Koter, A., \& Lamers, H. J. G. L.~M. 2001, Astron. Astrophys.
  v.369, p.574-588, 369, 61

\bibitem[{Vreugdenhil \& Koren(1993)}]{Vreugdenhil1993}
Vreugdenhil, C.~B. \& Koren, B. 1993 (Vieweg), 373

\bibitem[{Walder {et~al.}(2014)Walder, Melzani, Folini, Winisdoerffer, \&
  Favre}]{Walder2014}
Walder, R., Melzani, M., Folini, D., Winisdoerffer, C., \& Favre, J.~M. 2014,
  8th Int. Conf. Numer. Model. Sp. Plasma Flows (ASTRONUM 2013) ASP Conf. Ser.,
  488, 1

\bibitem[{Walter {et~al.}(2015)Walter, Lutovinov, Bozzo, \&
  Tsygankov}]{Walter15}
Walter, R., Lutovinov, A.~A., Bozzo, E., \& Tsygankov, S.~S. 2015, Astron.
  Astrophys. Rev., 23, 2

\bibitem[{Xia {et~al.}(2018)Xia, Teunissen, {El Mellah}, Chan{\'{e}}, \&
  Keppens}]{Xia2017}
Xia, C., Teunissen, J., {El Mellah}, I., Chan{\'{e}}, E., \& Keppens, R. 2018,
  Astrophys. J. Suppl. Ser., 234, 30

\bibitem[{Ziolkowski(1985)}]{Ziolkowski1985}
Ziolkowski, J. 1985, Acta Astron. (ISSN 0001-5237), vol. 35, no. 3-4, 1985, p.
  185-198., 35, 1

\end{thebibliography}
\end{tiny}

\end{document}